\documentclass[fleqn,usenatbib]{mnras}

\usepackage{newtxtext,newtxmath}

\usepackage[T1]{fontenc}

\DeclareRobustCommand{\VAN}[3]{#2}
\let\VANthebibliography\thebibliography
\def\thebibliography{\DeclareRobustCommand{\VAN}[3]{##3}\VANthebibliography}

\usepackage{graphicx}	
\usepackage{amsmath}	
\usepackage{pifont}
\usepackage{wasysym}

\usepackage{multirow}
\usepackage{dcolumn}
\newcolumntype{d}[1]{D{.}{.}{#1}}

\usepackage[]{xcolor}

\definecolor{myg}{cmyk}{0.75002,0,1,0}

\definecolor{msnote}{hsb:rgb}{0.492,0.492,0.492}

\newcommand{\flux}{$\rm erg\,s^{-1}cm^{-2}$}
\newcommand{\lum}{$\rm erg\,s^{-1}$}

\newcommand{\msun}{$\rm{M}_{\odot}$}

\newcommand{\aox}{$\alpha_{\rm ox}$}

\newcommand{\swift}{{\it Swift}}

\newcommand{\xmm}{{\it XMM-Newton}}

\newcommand{\nustar}{{\it NuSTAR}}
\newcommand{\fsoft}{$f_{\rm soft}$}
\newcommand{\fhard}{$f_{\rm hard}$}

\title[MWL variability of $\gamma$-NLS1 galaxies]{Multiwavelength variability of $\gamma$-ray emitting narrow-line Seyfert 1 galaxies}

\author[S. Yao \& S. Komossa]{
Su Yao\thanks{E-mail: syao@mpifr-bonn.mpg.de}
and 
S. Komossa\thanks{E-mail: astrokomossa@gmx.de}
\\
Max-Planck-Institut f\"ur Radioastronomie, Auf dem H{\"u}gel 69, 53121 Bonn, Germany\\
}

\date{Accepted XXX. Received YYY; in original form 2022}

\pubyear{2022}

\begin{document}
\label{firstpage}
\pagerange{\pageref{firstpage}--\pageref{lastpage}}
\maketitle

\begin{abstract}
As one of the drivers of feedback in active galactic nuclei (AGNs), the jets launched from supermassive black holes (SMBHs) are important for understanding the co-evolution of SMBHs and their host galaxies. 
However, the formation of AGN jets is far from clear. 
The discovery of $\gamma$-ray narrow-line Seyfert 1 (NLS1) galaxies during the past two decades has provided us with a  new means of studying the link between jets and accretion processes and the formation of jets. 
Here we explore the coupling of jet and accretion discs in seven bright $\gamma$-ray NLS1 galaxies by studying simultaneous optical/UV and X-ray observations of these systems taken by \swift. 
The results show that, except for 1H~0323+342 in which the X-rays are significantly contributed from the accretion disc, 
the observed X-ray emission of the other sources is dominated by the jet, and the accretion process makes little contribution if not absent. 
Although the origin of the X-ray emission is different,
the broad-band spectral shape characterized by \aox\ and the X-ray flux is found to follow the same evolutionary trend in 1H~0323+342, PMN~J0948+0022, 
and PKS~1502+036.
For the remaining sources, the trend is not observed 
or the sampling is not dense enough. 
\end{abstract}




\section{Introduction}

Powered by supermassive black holes (SMBHs, in the mass range 10$^6$--10$^{10}$\,\msun) at their centers,  
active galactic nuclei (AGNs) are among the most luminous long-lived sources in the universe. 
Collimated outflows of matter in form of jets, 
launched from the vicinity of the central accreting SMBH, are found in many AGNs. 
Jets and outflows are suggested to be one of the fundamental mechanisms by which the SMBH interacts with the host galaxy, regulating its evolution \citep{DiMatteo2005, Wagner2012}. 
However, the physical mechanism of jet formation and the role of black hole mass, magnetic fields and accretion disc properties are still not well understood \citep{Blandford2019}. 
The energetic particles in AGNs emit radiation at radio frequencies via the synchrotron process and make their jets detectable as luminous radio sources. 
But only less than 20 percent of all AGNs are found to be radio-loud \citep[][]{1995PASP..107..803U, 2002AJ....124.2364I}, 
while the rest are radio-quiet. The majority of the radio-quiet sources do have faint jets, and only few AGNs 
are not detected at all in the radio regime. However, the question is still open as to which parameters drive the most powerful, bright, relativistic jets in the radio-loud population in particular. 

The conventional picture established during the first few decades after the discovery of radio quasar was that
the relativistic AGN jets are preferentially associated with the heavier SMBHs residing in giant elliptical/bulge-dominated galaxies \citep[][]{1997ApJ...479..642B, 2003MNRAS.340.1095D, 2004MNRAS.355..196F}. 
This knowledge has been updated by the discoveries in the last two decades of samples of radio-loud narrow-line Seyfert 1 (NLS1) galaxies and the more massive radio-loud narrow-line type 1 quasars \citep{2006AJ....132..531K} that showed SMBH masses below 10$^8$ M$_{\odot}$, below the classical radio-loud regime.
The vast majority of the latter still lacks host images, but among the very small number of them that do
have host imaging, a few disc-like galaxies 
have been identified \citep[e.g.][]{2007ApJ...658L..13Z, 
2018A&A...619A..69J, 2020MNRAS.492.1450O}.

In particular, some radio-loud NLS1 galaxies show persistent or flaring $\gamma$-ray emission independently re-confirming their launch of powerful relativistic jets
\citep[e.g.][]{2009ApJ...699..976A, 2009ApJ...707L.142A}. The presence of such jets was already earlier found from radio observations
\citep[e.g.,][]{2006AJ....132..531K, 2008ApJ...685..801Y}. 
NLS1 galaxies generally host less massive but rapidly growing SMBHs 
and NLS1s galaxies are suggested to be located at the end of the AGN parameter space opposite to the classical blazars \citep[][]{2002ApJ...565...78B}.
Thus, NLS1 galaxies  provide us with new laboratories for studying open questions related to the formation and evolution of relativistic jets in AGNs.  

For this study, we selected bright radio-loud, gamma-ray detected NLS1 galaxies with multiyear coverage obtained by the Neil Gehrels \swift\ observatory \citep[\swift\ hereafter;][]{Gehrels2004}. 
The targets of our study (Table~\ref{tab:gnls1}) 
have been studied in selected time intervals and/or wavebands before 
\citep[e.g.,][]{2010ApJ...715L.113L, 2011MNRAS.413.2365C, 2012ApJ...759L..31J, 2012A&A...548A.106F, 2013ApJ...773...85E, 2013ApJ...775L..26I, 2015AJ....150...23Y, 2015MNRAS.454L..16Y, 
Foschini2015, 
2018rnls.confE..42G, 2018A&A...618A..92A, 
2018MNRAS.477.5127Y, 2019MNRAS.487L..40Y, 2020MNRAS.496.2213D, 2020MNRAS.498..859D, 2021A&A...654A.125B, 2021ApJS..255...10M, 2022MNRAS.tmp.1571O}. 
Here, we combine archival \swift\ data and our own \swift\ data until  2021 December. 
Focus is on recent data since 2019, but we also add pre-2019 data to the analysis, for comparison purposes and in order to identify systematic trends.  
In particular, we address
the broad-band spectral evolution using the simultaneous optical/UV and X-ray observations obtained by the X-ray telescope \citep[XRT;][]{2005SSRv..120..165B} 
and 
ultra-violet--optical telescope 
\citep[UVOT;][]{2005SSRv..120...95R} 
onboard \swift. 
We note that several of the sources in our study are formally quasars, however, for simplicity we continue to refer to all of them as "Seyfert" galaxies hereafter, as is a common habit when discussing NLS1 galaxies. 

This paper is structured as follows: 
In Section~\ref{sec:obs} we describe the data reduction for the XRT and UVOT, respectively. 
The results and the analysis based on the optical/UV and X-ray data are provided in Section~\ref{sec:results}. 
In Section~\ref{sec:discussion} we discuss the results of the \swift\ observations and finally a brief summary is given in Section~\ref{sec:summary}. 
Throughout this work, we use a cosmology with $H_{\rm 0}$=70 km\,s$^{-1}$\,Mpc$^{-1}$, $\Omega_{\rm M}$=0.3 and $\Omega_{\rm \Lambda}$=0.7.

\newcounter{tabref}
\begin{table*}
    \caption{Summary of the $\gamma$-ray NLS1 galaxies in our study. }
    \label{tab:gnls1}
    \begin{center}
    \begin{tabular}{lcccccccclccclc}
        \hline
        Name & R.A.  &  Dec.  &  $z$  & $\gamma$-ray detection & BH mass & BH mass \\
        & (J2000) & (J2000) & & Ref. & [10$^{7}$\,\msun] & Ref. \\
        (1)  &  (2)  &  (3)  &  (4)  &  (5)  &  (6)  &  (7) \\%
        \hline
        1H~0323+342  & 03:24:41.16 & +34:10:45.9 & 0.063 &  \ref{Abdo20092} & ${3.4}$ & \ref{WF2016} & \\ %
        SDSS~J094635.06+101706.1 & 09:46:35.07 & +10:17:06.1 & 1.004 & \ref{Abdo2010} & 19 & \ref{YS2019} \\ %
        PMN~J0948+0022 & 09:48:57.32 & +00:22:25.6 & 0.584 & \ref{Abdo20091} & 4 & \ref{ZHY2003} \\ %
        SDSS~J122222.55+041315.7 & 12:22:22.55 & +04:13:15.8 & 0.966 & \ref{Abdo2010},\ref{YS2015} & 20 & \ref{YS2015} \\ %
        PKS~1502+036 & 15:05:06.48 & +03:26:30.8 & 0.408 & \ref{Abdo20092} & 0.4 & \ref{YWM2008} \\ %
        PKS~2004$-$447 & 20:07:55.18 & $-$44:34:44.2 & 0.24 & \ref{Abdo20092} & 7 & \ref{Gokus2021} \\ %
        SDSS~J211852.96$-$073227.5 & 21:18:52.96 & $-$07:32:27.6 & 0.260 & \ref{YH2018},\ref{Paliya2018} & 3.4 & \ref{YH2018} \\ %
        \hline
    \end{tabular}
    \parbox[]{\textwidth}{
    {\it Notes.} 
    Column (1): source name. 
    Column (2) and (3): right ascension (R.A.) and declination (Dec.) of the galaxy. 
    Column (4): redshift. 
    Column (5): References for the $\gamma$-ray detection of the source.
    Column (6) and (7): black hole mass and reference. 
    {References:} 
    \refstepcounter{tabref}\label{Abdo20092}(\ref{Abdo20092}) \citet{2009ApJ...707L.142A}; 
    \refstepcounter{tabref}\label{Abdo2010}(\ref{Abdo2010})\citet{2010ApJS..188..405A}; 
    \refstepcounter{tabref}\label{Abdo20091}(\ref{Abdo20091}) \citet{2009ApJ...699..976A}; 
    \refstepcounter{tabref}\label{YS2015}(\ref{YS2015}) \citet{2015MNRAS.454L..16Y}; 
    \refstepcounter{tabref}\label{YH2018}(\ref{YH2018}) \citet{2018MNRAS.477.5127Y}; 
    \refstepcounter{tabref}\label{Paliya2018}(\ref{Paliya2018})\citet{2018ApJ...853L...2P}; 
    \refstepcounter{tabref}\label{WF2016}(\ref{WF2016}) \citet{2016ApJ...824..149W}; 
    \refstepcounter{tabref}\label{YS2019}(\ref{YS2019}) \citet{2019MNRAS.487L..40Y}; 
    \refstepcounter{tabref}\label{ZHY2003}(\ref{ZHY2003}) \citet{2003ApJ...584..147Z}; 
    \refstepcounter{tabref}\label{YWM2008}(\ref{YWM2008}) \citet{2008ApJ...685..801Y}; 
    \refstepcounter{tabref}\label{Gokus2021}(\ref{Gokus2021}) \citet{2021A&A...649A..77G}; 
    }
    \end{center}
\end{table*}

\section{\swift\ Observations and data reduction}
\label{sec:obs}

\subsection{XRT}

\subsubsection{Data reduction}

Data taken with the \swift/XRT 
are reprocessed following standard procedures using the task \textit{xrtpipeline}. 
The photon counting (PC) mode \citep{2004SPIE.5165..217H} data are selected for the analysis. 
The images are extracted first and visually inspected to exclude those observations in which the source is at the edge or outside of the field of view. 
To obtain the spectra, 
the source counts are extracted from a circle of 50\arcsec\ radius centered on the source position 
and the background is estimated from an annulus with inner radius of 60\arcsec\ and three times the area of the source region, 
except for 1H~0323+342 which sometimes has brightened to a count rate higher than 0.5 counts\,s$^{-1}$. These data are affected by pile-up. 
In order to determine at which point the data and the XRT point spread function (PSF) model diverge, 
the XRT image from the observation in which 1H~0323+342 has highest count rate is modeled by a King function 
$\text{PSF}(r) = \left[ 1+(r/r_{\rm c})^{2} \right]^{\beta}$, 
where $r_{\rm c}=5.8$, $\beta=1.55$ and $r$ is the distance to the PSF center. 
The King function is fitted to the outer wings of the PSF at distance larger than 15\arcsec. 
The best-fit model is then visually compared with the data, 
and the diverge appears below $\sim$8\arcsec. 
Thus, for all the XRT observations on 1H~0323+342 with count rate $>0.5$\,counts\,s$^{-1}$ 
the source events are extracted from an annulus with inner/outer radius of 8\arcsec/60\arcsec, 
and the backgrounds are estimated from an annulus with inner/outer radius of 70\arcsec/90\arcsec.

The observations in which the sources are not detected ($<5\sigma$) are excluded. 
Full grade selection (0--12) is adopted to extract the source and background spectra. 
The 
ancillary response files 
are built using task \textit{xrtmkarf} with the corresponding exposure maps to correct the PSF if any. 
The relevant 
response matrix files 
given in the output of \textit{xrtmkarf} are used. 
The spectra with more than 125 counts are grouped such that they have at least 25 counts in each bin,
and the $\chi^{2}$ minimization is used to fit model to the spectra. 
For the spectra with 125 or less counts, 
the spectra are grouped such that each spectrum has 5 to 10 bins, 
and the \textit{Cash} statistic is adopted to fit the spectra. 
The software package XSPEC \citep{1996ASPC..101...17A} was used for
spectral analysis.
A summary of the X-ray observations used in our analysis is given in Table~\ref{tab:observations}. 

\subsubsection{Spectral fitting and flux estimation}

\begin{figure*}
  \centering
  \includegraphics[width=0.99\textwidth]{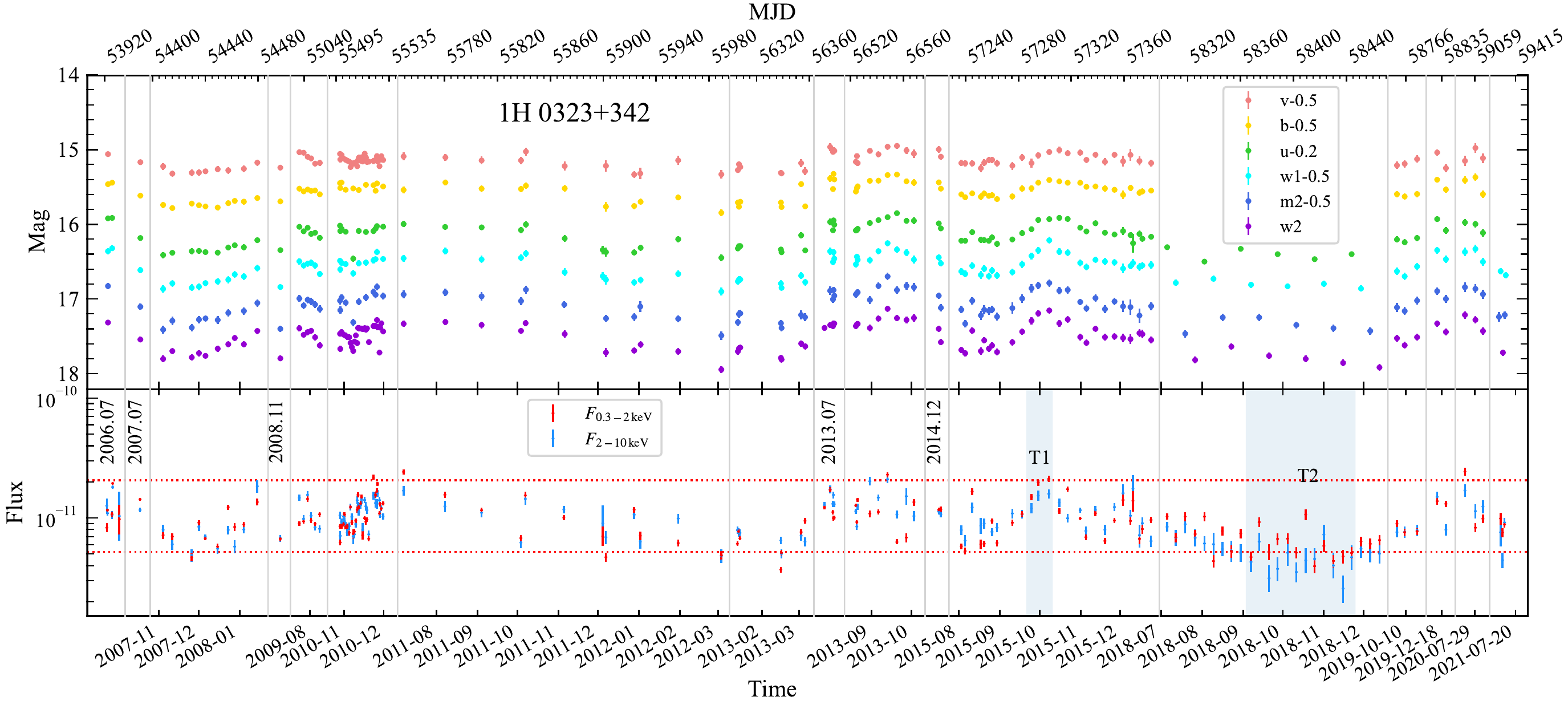}
  \caption{
  \swift\ light curve of 1H~0323+342. 
  The time along the x-axis is not continuous. 
  Time intervals without any data were removed for visual clarity. 
  The vertical grey lines in the figure mark the boundaries of each time windows in which there exist observed data.
  {\it Upper panel:}
  light curve in the $v$, $b$, $u$, $w1$, $m2$ and $w2$ bands in units of magnitude. 
  The magnitudes of the source are offset for visual clarity 
  and are not corrected for Galactic extinction. 
  {\it Lower panel:}
  The X-ray flux is reported in the 0.3--2\,keV band (red points) and the 2--10\,keV band (blue points). 
  The red dotted lines mark the higher and lower flux thresholds, 
  above and below which the high and low states are selected based on 0.3--2\,keV X-ray fluxes.  
  The light blue shaded areas mark the epochs selected for X-ray spectral stacking (see Section~\ref{sec:stacking}). 
  }
  \label{fig:lc_1h0323}
\end{figure*}

\begin{figure*}
  \centering
  \includegraphics[width=0.98\textwidth]{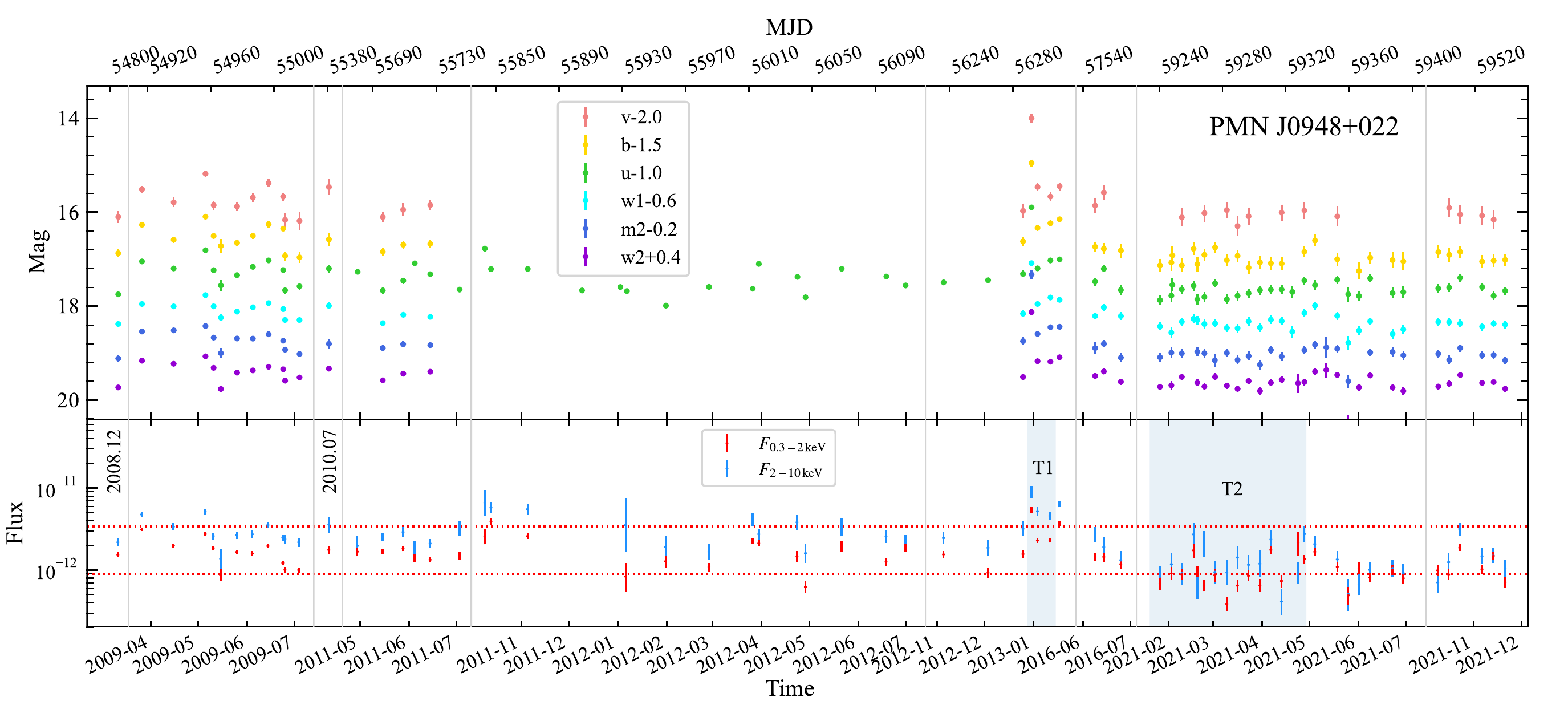}
  \caption{
  Same as Figure~\ref{fig:lc_1h0323} but for PMN~J0948+0022. 
  }
  \label{fig:lc_0948}
\end{figure*}

\begin{figure*}
  \centering
  \includegraphics[width=0.98\textwidth]{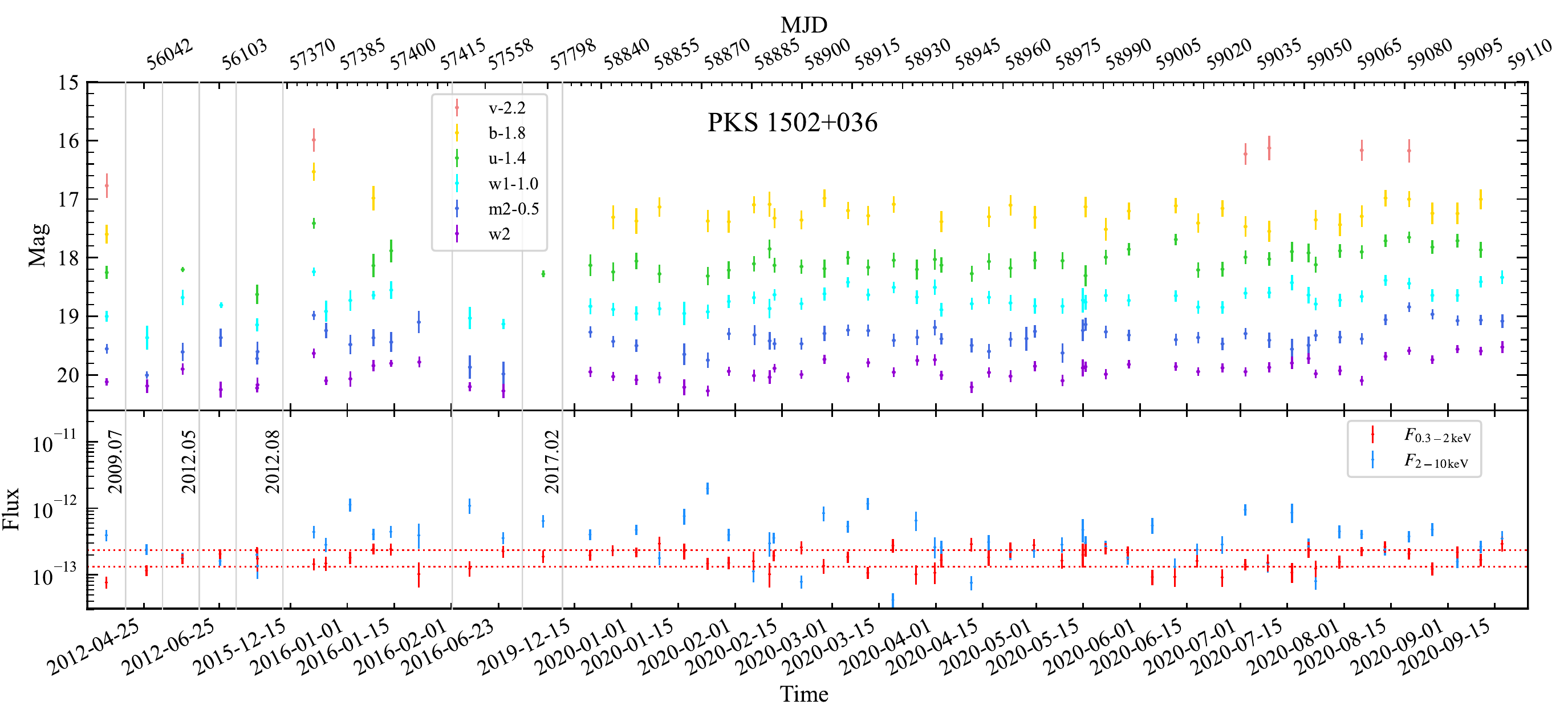}
  \caption{
  Same as Figure~\ref{fig:lc_1h0323} but for PKS~1502+036. 
  }
  \label{fig:lc_1502}
\end{figure*}

\begin{figure*}
  \centering
  \includegraphics[width=0.98\textwidth]{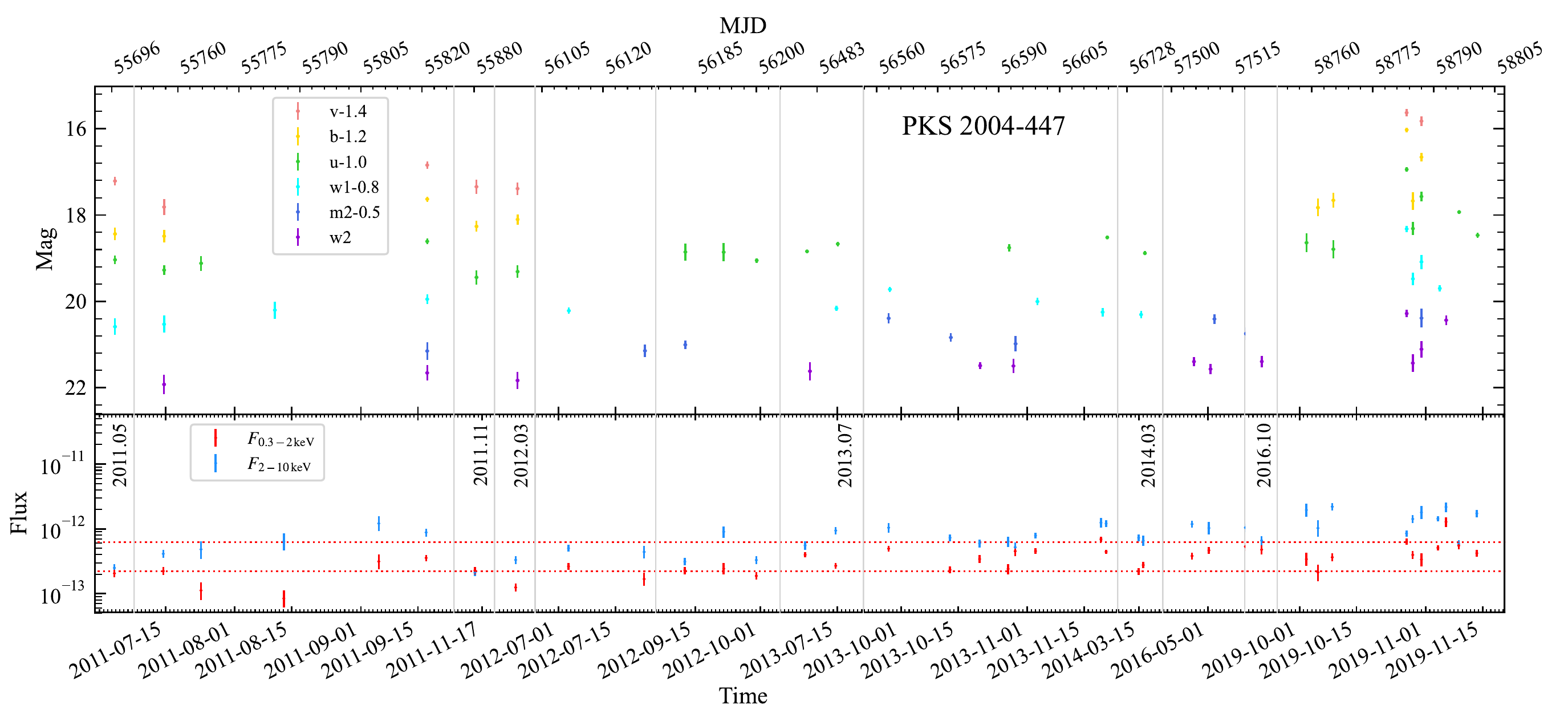}
  \caption{
  Same as Figure~\ref{fig:lc_1h0323} but for PKS~2004-447. 
  }
  \label{fig:lc_2004}
\end{figure*}

\begin{figure}
  \centering
  \includegraphics[width=0.99\columnwidth]{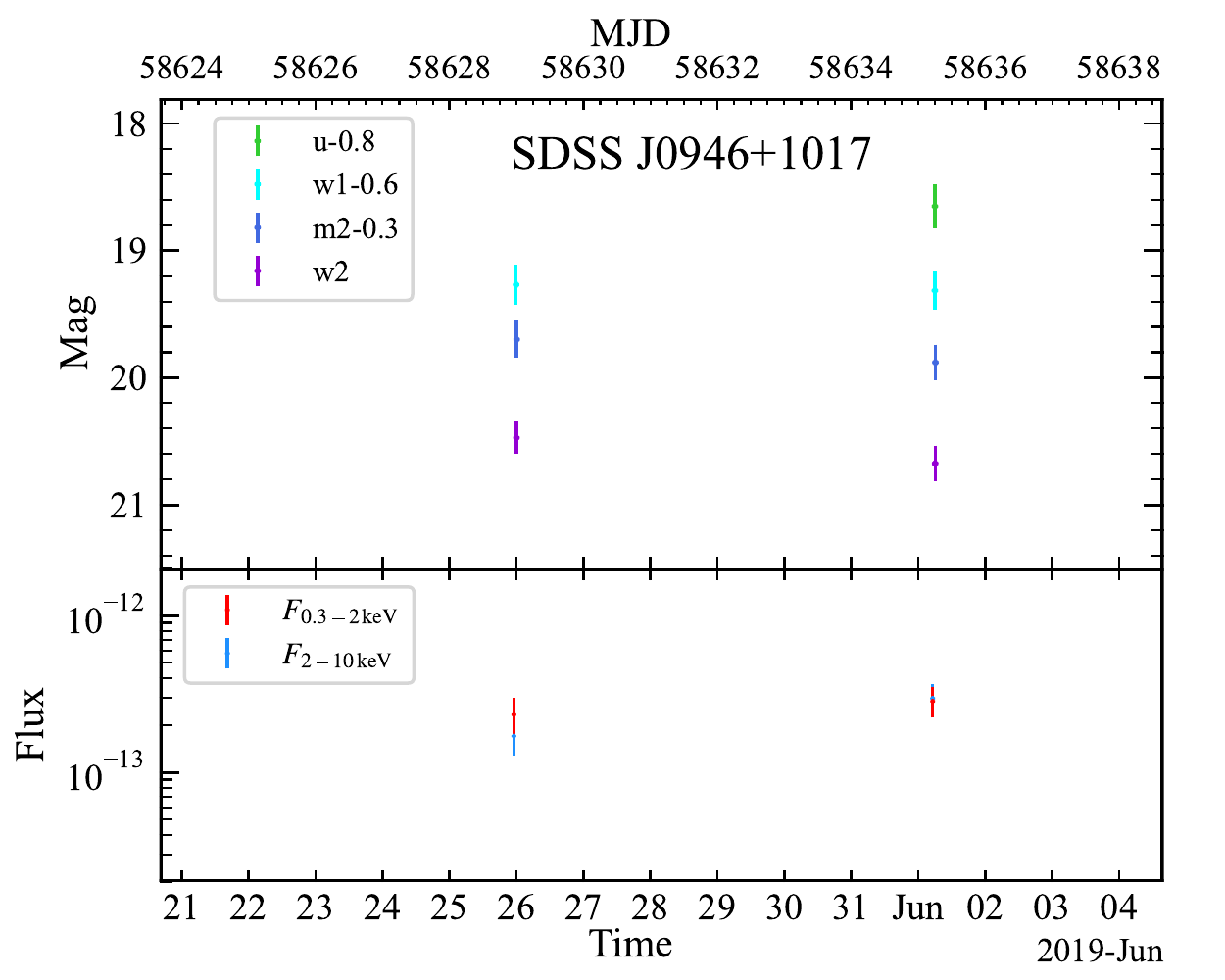}
  \caption{
  Light curve of J0946+1017. 
  {\it Upper panel:}
  light curve in the $u$, $w1$, $m2$ and $w2$ bands in units of magnitude. 
  The magnitudes of the source are offset for visual clarity 
  and are not corrected for extinction. 
  {\it Lower panel:}
  The X-ray flux in 0.3--2\,keV (red) and 2--10\,keV (blue). 
  }
  \label{fig:lc_0946}
\end{figure}

\begin{figure}
  \centering
  \includegraphics[width=0.99\columnwidth]{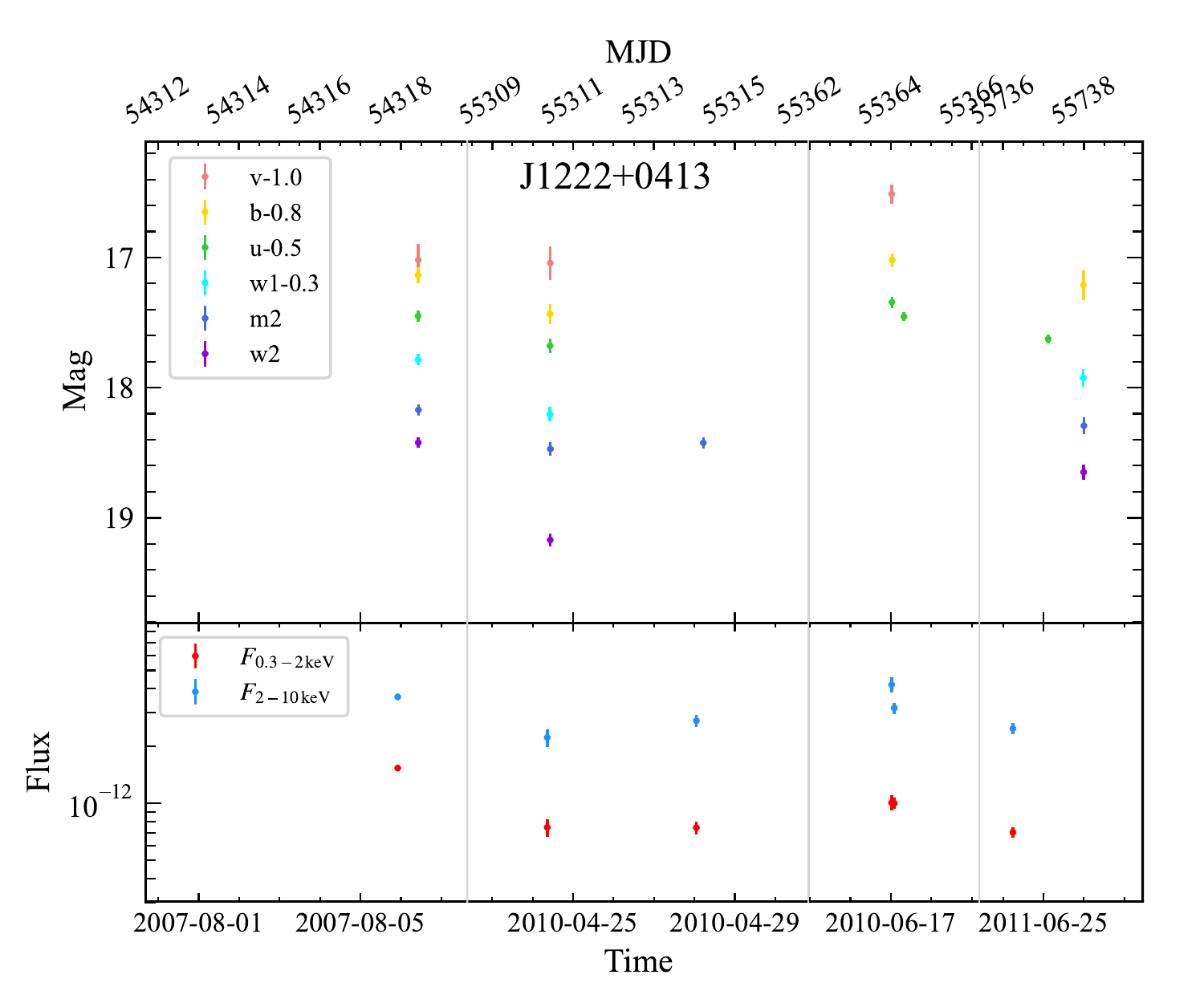}
  \caption{
  Light curve of SDSS~J1222+0413. 
  {\it Upper panel:}
  light curve in the $v$, $b$, $u$, $w1$, $m2$ and $w2$ bands in units of magnitude. 
  The magnitudes of the source are offset for visual clarity 
  and are not corrected for extinction. 
  {\it Lower panel:}
  The X-ray flux light curve in 0.3--2\,keV and 2--10\,keV. 
  }
  \label{fig:lc_1222}
\end{figure}

\begin{figure*}
  \centering
  \includegraphics[height=0.40\textwidth]{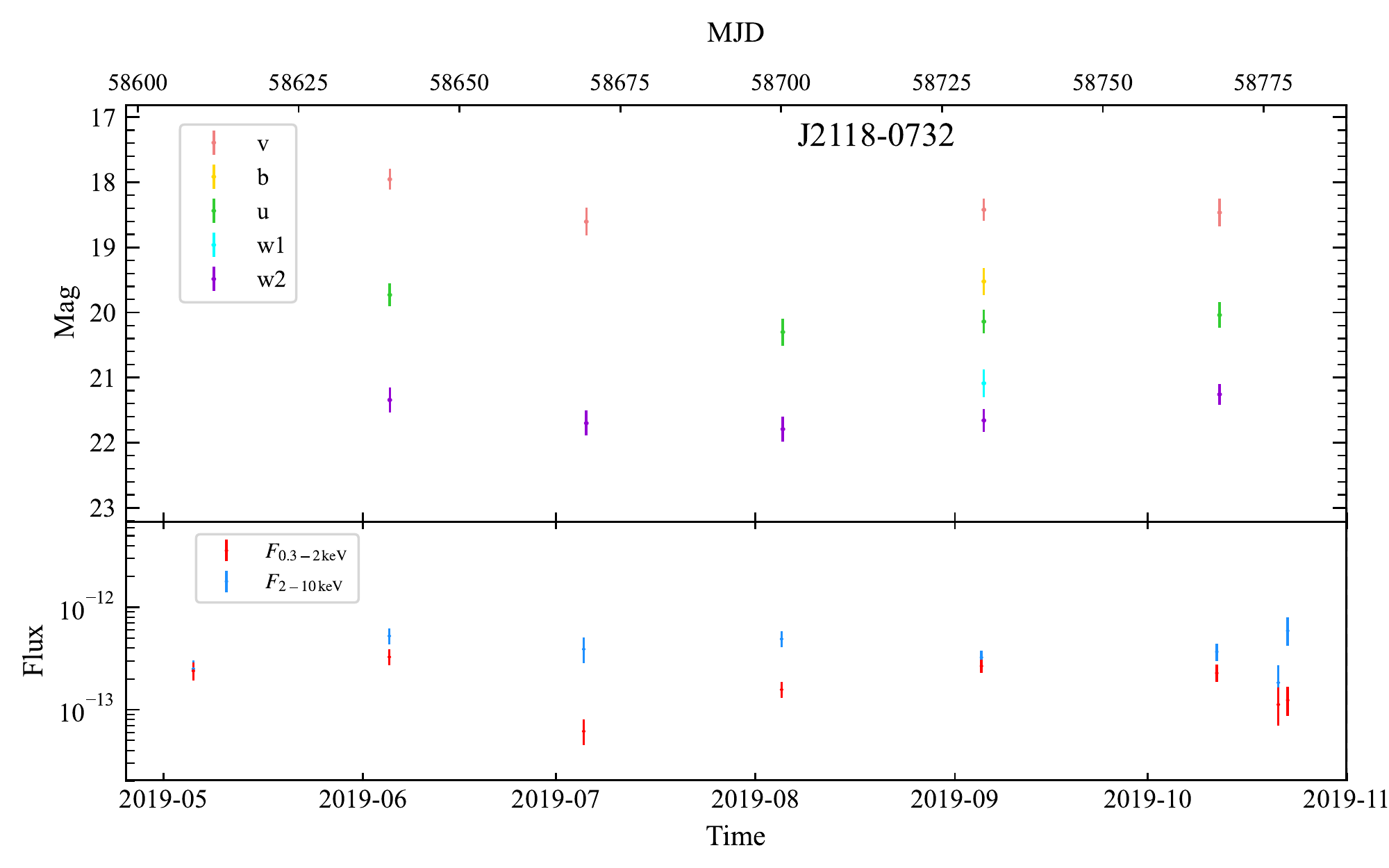}
  \caption{
  Light curve of SDSS~J2118$-$0732. 
  {\it Upper panel:}
  light curve in the $v$, $b$, $u$, $w1$ and $w2$ bands in units of magnitude. 
  The magnitudes of the source 
  are not corrected for extinction. 
  {\it Lower panel:}
  The X-ray flux light curve in 0.3--2\,keV and 2--10\,keV. 
  }
  \label{fig:lc_j2118}
\end{figure*}

All the spectra with 125 or less counts, 
and most of the spectra with more than 125 counts can be well fitted by a single power law 
with Galactic neutral hydrogen absorption, 
except for some spectra of 1H~0323+342 when the $\chi^{2}$-minimization gives a very poor, unacceptable fit, 
indicating an extra component other than a single power law is required. 
The spectra revealing a $\chi^{2}$ higher than the 90\% quantile of the $\chi^{2}$ distribution given their degrees of freedom (dof) are then selected and fitted with a blackbody component plus a single power-law model. 
The absorption is fixed at the Galactic value and 
the blackbody temperature is forced to be in the range of $0.03\,\text{keV}<\text{Tk}<0.2\,\text{keV}$ during the fitting. 
The absorption-corrected flux is estimated by \textit{cflux}.
The fluxes in the 0.3--2\,keV band (\fsoft) and 2--10\,keV (\fhard) band of the sources are shown in Figure~\ref{fig:lc_1h0323}--\ref{fig:lc_j2118}.
A summary of the results obtained from XRT observations is given in Table~\ref{tab:results}.

\begin{table*}
    \caption{Summary of the \swift\ observations. }
    \label{tab:observations}
    \begin{center}
    \begin{tabular}{llclccccccclccclc}
        \hline
        \multirow{2}{*}{Name} & 
        \multicolumn{1}{c}{Time} & 
        \multicolumn{2}{c}{X-ray} & 
        \multicolumn{6}{c}{$N_{\rm obs}$ of optical/UV} & 
        \\
        & 
        \multicolumn{1}{c}{Coverage} & 
        \multicolumn{1}{c}{$N_{\rm obs}$} & 
        \multicolumn{1}{c}{Exp. (s)} & 
        \multicolumn{1}{c}{$v$} & 
        \multicolumn{1}{c}{$b$} & 
        \multicolumn{1}{c}{$u$} & 
        \multicolumn{1}{c}{$w1$} & 
        \multicolumn{1}{c}{$m2$} & 
        \multicolumn{1}{c}{$w2$} & 
        \\
        (1) & (2) & (3) & (4) & (5) & (6) & (7) & (8) & (9) & (10) 
        \\%
        \hline
        1H~0323+342 & 
            2006 Jul 05 -- 2021 Jul 11 & 149 & 74 -- 8821 & 113 & 97 & 105 & 107 & 102 & 123 
            \\
        SDSS~J094635.06+101706.1 & 
            2019 May 25 -- 2019 Jun 01 & 2 & 1743 -- 1875 & 0 & 0 & 1 & 2 & 2 & 2 
            \\
        PMN~J0948+0022 & 
            2008 Dec 05 -- 2021 Nov 20 & 70 & 187 -- 7754 & 34 & 51 & 73 & 52 & 50 & 53 \\
        SDSS~J122222.55+041315.7 & 
            2007 Aug 05 -- 2011 Jun 24 & 6 & 1561 -- 9178 & 3 & 4 & 5 & 3 & 4 & 3 
            \\
        PKS~1502+036 & 
            2009 Jul 25 -- 2020 Sep 17 & 59 & 631 -- 5157 & 6 & 36 & 47 & 55 & 57 & 58 
            \\
        PKS~2004$-$447 & 
            2011 May 15 -- 2019 Nov 13 & 39 & 714 -- 12\,180 & 7 & 10 & 21 & 14 & 9 & 13 
            \\ 
        SDSS~J211852.96$-$073227.5 & 
            2019 May 05 -- 2019 Oct 22 & 8 & 1204 -- 3771 & 4 & 1 & 4 & 1 & 0 & 5 
        \smallskip
        \\
        \hline
    \end{tabular}
    \parbox[]{\textwidth}{
    {\it Notes.} 
    Column (1): source name. 
    Column (2): time coverage of the \swift\ observations. 
    Column (3) and (4): number of the X-ray observations and the range of exposure times of the X-ray observations. 
    Column (5)--(10): number of the optical/UV observations in $v$, $b$, $u$, $w1$,  $m2$ and $w2$ band, respectively. We note that $N_{\rm obs}$ refers to the number of observations used in the following analysis, excluding observations with no detection. 
    }
    \end{center}
\end{table*}

\subsection{UVOT}

The NLS1 galaxies of this study have also been observed with the \swift/UVOT, 
in up to all three optical 
and up to all three UV bands 
[with filters: $v$ (5468\AA), $b$ (4392\AA), $u$ (3465\AA),  $w1$ (2600\AA), $m2$ (2246\AA), 
and $w2$ (1928\AA), where values in brackets represent the filter central wavelengths \citep{2008MNRAS.383..627P}]. 
Some sources have only been observed in selected bands. 
For each observation, 
data in each filter are co-added using the task \textit{uvotimsum} after aspect correction. 
The images are visually inspected to exclude those in which the source is on the edges or out of the field of view. 
The source counts in each available filter were extracted in a circle with radius of 5\arcsec,
while the background was selected in a nearby source-free region of radius 15\arcsec.
The background-corrected counts are then converted into fluxes based on the latest calibration as described by \citet{2008MNRAS.383..627P} and \citet{2010MNRAS.406.1687B} using the task \textit{uvotsource}. 
A  summary of the UVOT observations and results are given in Table~\ref{tab:observations} and Table~\ref{tab:results}, respectively.
The total duration of each UVOT observation is the same as the corresponding XRT duration. 
Each UVOT band $v$:$b$:$u$:$w1$:$m2$:$w2$ 
is observed with a ratio of 1:1:1:2:3:4 of the total exposure time, respectively.
Correction of the UVOT fluxes for Galactic reddening towards the individual NLS1 galaxies was carried out using the values of $E_{\rm{(B-V)}}$ from \citet{2011ApJ...737..103S} and the reddening curves of \citet{1999PASP..111...63F}. 
We use the extinction-corrected optical/UV flux in the following analysis unless mentioned otherwise.

\begin{table*}
\renewcommand{\arraystretch}{1.2}
    \caption{\swift\ results. }
    \label{tab:results}
    \begin{center}
    \begin{tabular}{lccccccccclccclc}
        \hline
        \multirow{2}{*}{Name} & 
        \multicolumn{4}{c}{Variability} & 
        \multirow{2}{*}{$\Gamma$} &
        \multicolumn{2}{c}{$L_{\rm X,peak}$} & 
        \multirow{2}{*}{$\alpha_{\rm UV}$} & 
        \multirow{2}{*}{\aox} &\\
        & optical & UV & X$_{\rm soft}$ & X$_{\rm hard}$ & & soft & hard \\
        & & & & & & \multicolumn{2}{c}{[$10^{44}$\,\lum]} \\
        (1)  &  (2)  &  (3)  &  (4)  &  (5)  &  (6)  &  (7)  &  (8)  & (9) & (10) \\%
        \hline
        1H~0323+342 & 1.4 & 2.1 & 6.6 & 8.2 & (1.1, 2.4) & 2.3 & 2.0 & ($-0.34$, 0.01) & (1.14, 1.38) \\
        SDSS~J094635.06+101706.1 & $\dots$ & 1.2 & 1.2 & 1.7 & (1.9, 2.1) & 15.6 & 16.2 & (1.26, 1.74) & (1.20, 1.28) \\
        PMN~J0948+0022 & $>$8.5 & 15.1 & 14.0 & 21.9 & (1.1, 2.3) & 77.5 & 130.2 & ($-0.40$, 0.02) & (1.02, 1.30) \\ 
        SDSS~J122222.55+041315.7 & 1.6 & 2.0 & 2.2 & 1.9 & (1.0, 1.4) & 76.1 & 209.6 & ($-0.42$, 0.08) & (1.13, 1.21) \\ 
        PKS~1502+036 & 2.1 & 2.0 & 3.8 & 47.4 & (0.4, 2.9) & 1.8 & 11.9 & ($-0.49$, 0.06) & (1.15, 1.48) \\ 
        PKS~2004$-$447 & 7.5 & $>$5.6 & 15.2 & 10.3 & (0.7, 1.9) & 2.2 & 3.8 & (1.39, 2.98) & (0.84, 1.17) \\ 
        SDSS~J211852.96$-$073227.5 & $>$1.9 & 1.6 & 5.3 & 3.2 & (0.8, 1.9) & 0.7 & 1.2 & (0.73, 1.25) & 1.20 
        \smallskip
        \\
        \hline
    \end{tabular}
    \parbox[]{\textwidth}{
    {\it Notes.} 
    Column (1): source name. 
    Column (2) to (5): factor of flux variation in the optical, UV,  soft X-rays (0.3--2\,keV) and hard X-rays (2--10\,keV), respectively.
    A lower limit on the variability in the optical or UV is obtained when a flux upper limit given by {\it uvotsource} is lower than all the detections. 
    Column (6): X-ray photon index range from spectral fitting. 
    Column (7) and (8): {\em{isotropic}} X-ray  peak luminosity in the soft and hard X-ray band, respectively, in units of $10^{44}$\,\lum. 
    Column (9): spectral index between optical and UV, $\alpha_{\rm uv}$, where $f_{\nu}\propto\nu^{-\alpha_{\rm uv}}$, calculated using the $v$ band and the $w2$ band. 
    When there is no or too few $v$ or $w2$ data, a nearby band is used (see Section~\ref{sec:swsed}). 
    Column (10): \aox\ calculated using the extinction corrected flux density at the $w1$ band and at 2\,keV. 
    }
    \end{center}
\end{table*}

        %

\section{Results}
\label{sec:results}

\subsection{Broadband spectral slope \aox}

To quantify the broadband spectral slope, 
the spectral index \aox\ between optical/UV and X-ray is calculated as
$\alpha_{\rm ox}=-0.384\log\left(f_{2500}/f_{2\rm\,keV}\right)$,
where $f_{2500}$ and $f_{2\rm\,keV}$ are the flux densities at 2500\,\AA\ and 2\,keV, respectively. 
Here the flux density obtained from $w1$ band centered at $\sim2600$\,\AA\ is adopted. 
The flux density at 2\,keV is obtained from the best-fit model to each XRT spectrum.

\begin{figure*}
\centering
	\begin{tabular}{cc}
		\includegraphics[width=0.96\columnwidth]{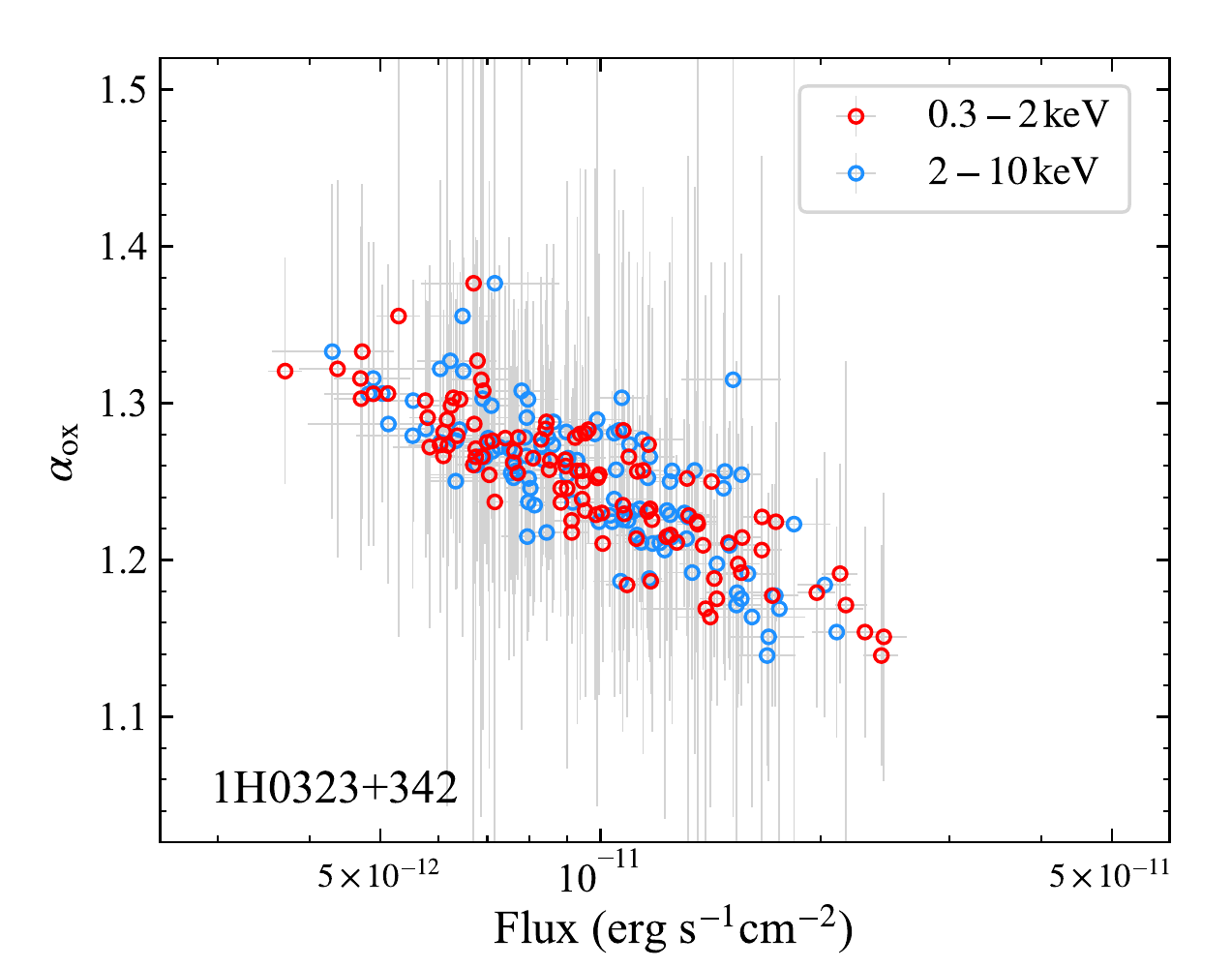} &
		\includegraphics[width=0.96\columnwidth]{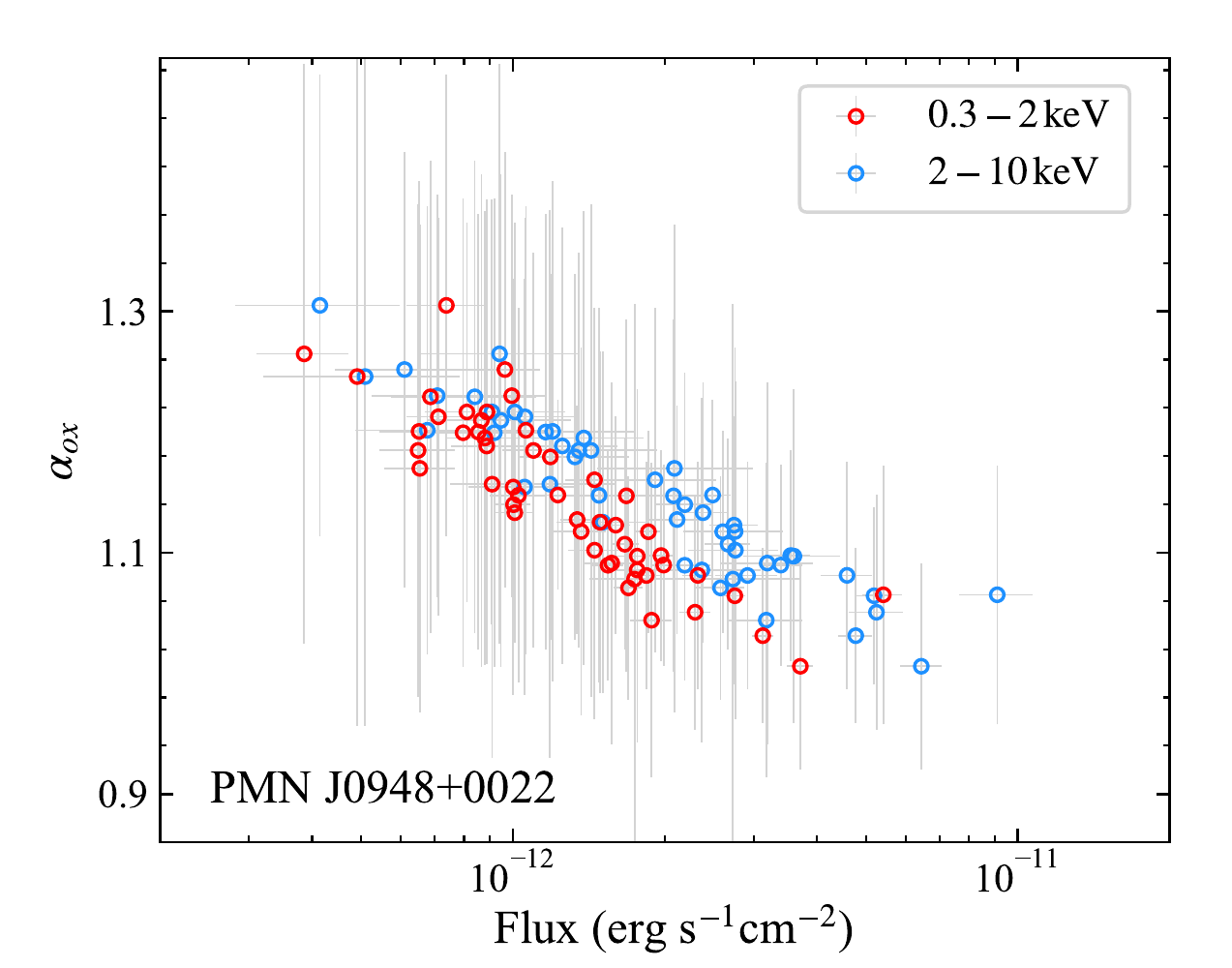} \\
		\includegraphics[width=0.96\columnwidth]{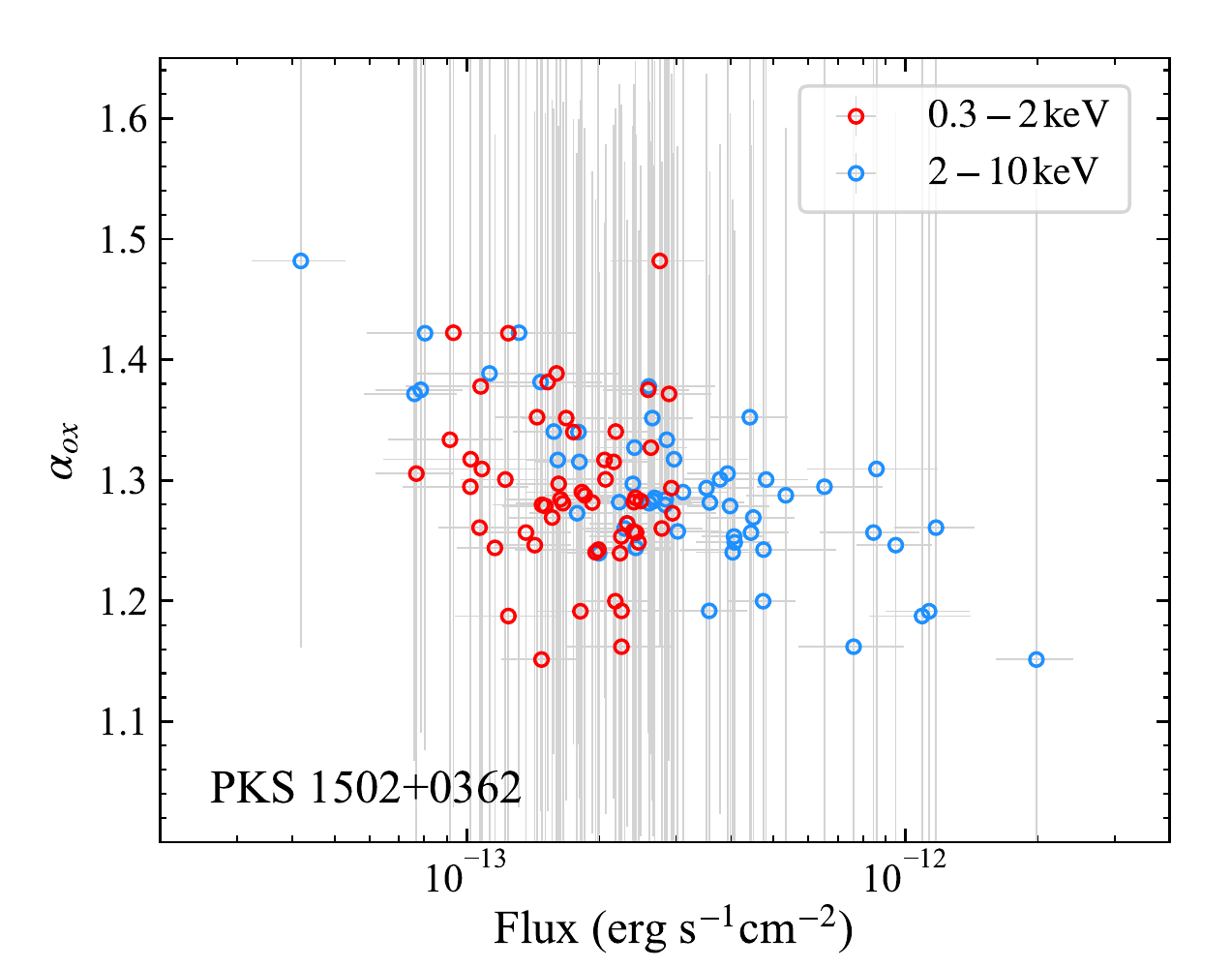} &
		\includegraphics[width=0.96\columnwidth]{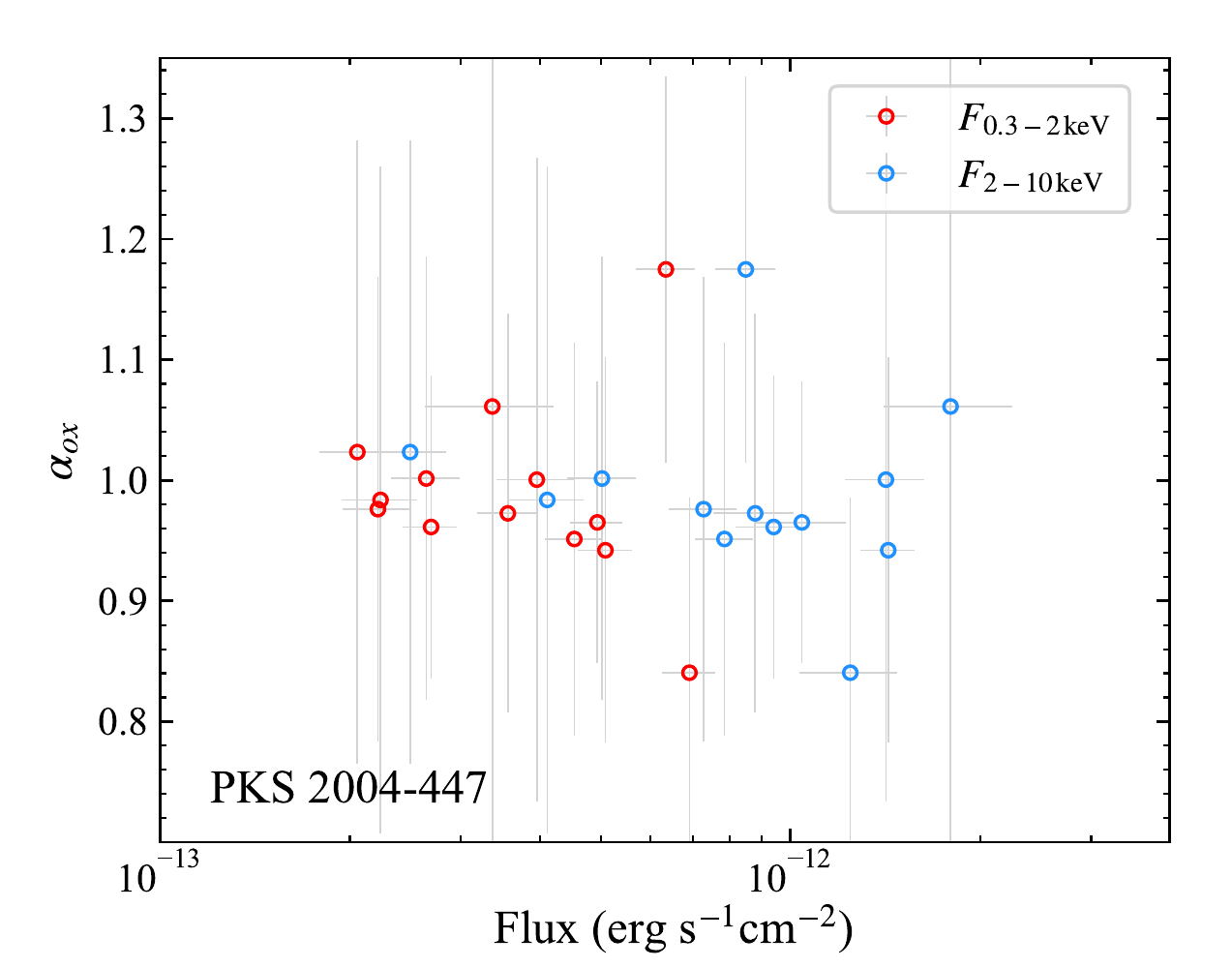} \\
	\end{tabular}
	\caption{%
	The \aox\ versus the X-ray flux for 1H~0323+342 (upper left), 
	PMN~J0948+0022 (upper right), 
	PKS~1502+036 (lower left) 
	and 
	PKS~2004-447 (lower right). 
	The X-ray flux in 0.3--2\,keV and 2--10\,keV are colored by red and blue, respectively. 
	The error bars are colored in grey. 
	}
\label{fig:aox}
\end{figure*}

To explore the broadband spectral slope variation along with the flux, 
we show the \aox\ vs. flux for 1H~0323+342, PMN~J0948+0022, PKS~1502+0362 and PKS~2004-447 in Figure~\ref{fig:aox}. 
The other three sources are not shown because they were observed only for a few times simultaneously in $w1$ and X-rays. 
As can be seen, 
although with large uncertainties there is a clear trend of softer \aox\ with decreasing X-rays flux in 1H~0323+342 and PMN~J0948+0022. 
In PKS~1502+0362, this trend is still obvious for the hard X-rays, but vanishes for the X-rays below 2\,keV. 
In PKS~2004-447, there is no clear trend for both soft and hard X-rays. 

\subsection{UV--X-ray SEDs}
\label{sec:swsed}

\begin{figure*}
\centering
	\begin{tabular}{cc}
		\includegraphics[width=0.80\columnwidth]{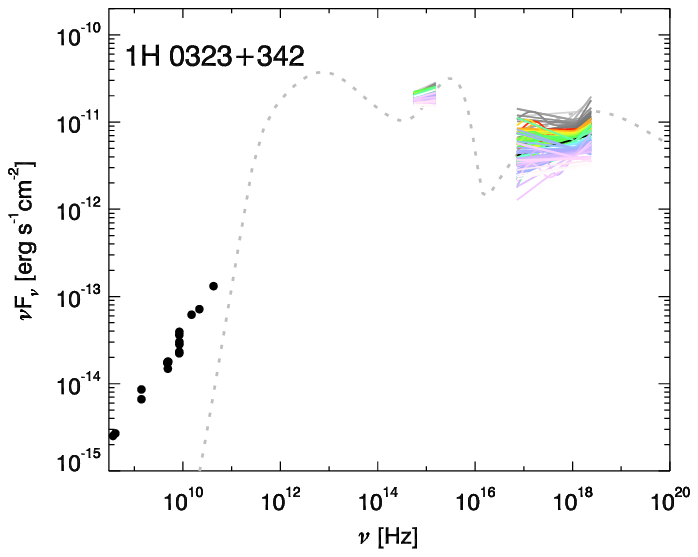} & 
		\includegraphics[width=0.80\columnwidth]{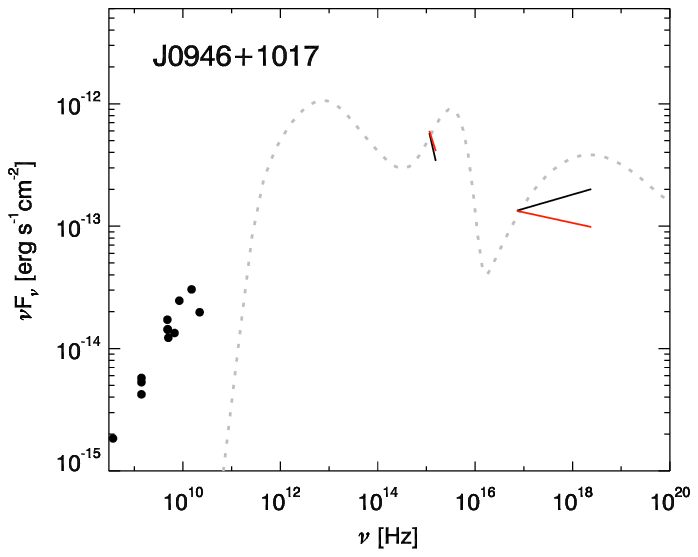} \\
		\includegraphics[width=0.80\columnwidth]{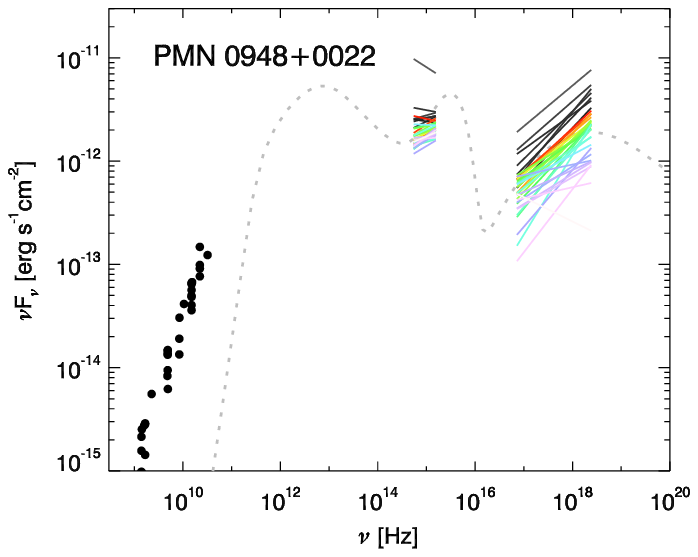} &
		\includegraphics[width=0.80\columnwidth]{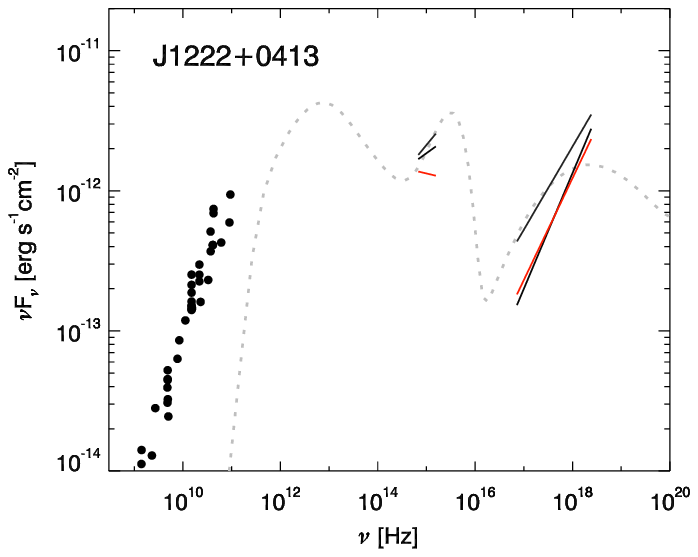} \\ 
		\includegraphics[width=0.80\columnwidth]{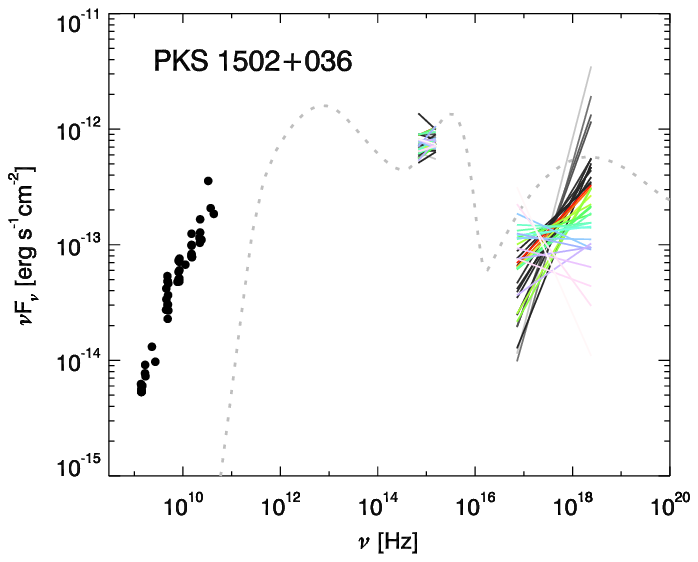} &
		\includegraphics[width=0.80\columnwidth]{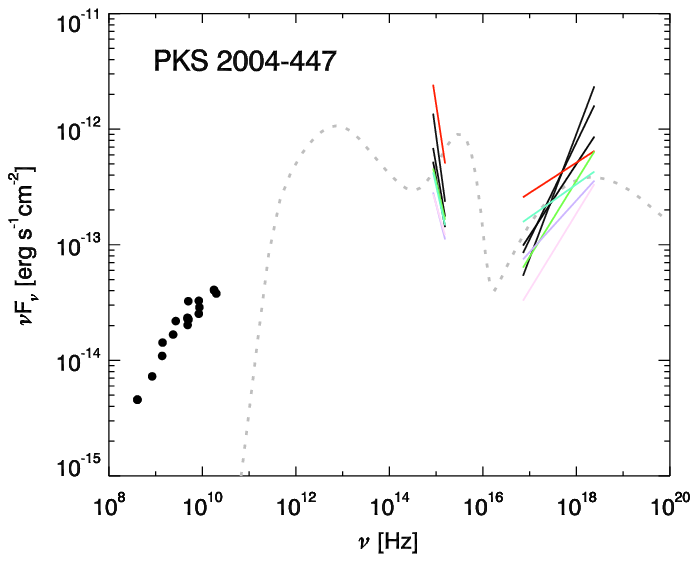} \\
		\includegraphics[width=0.80\columnwidth]{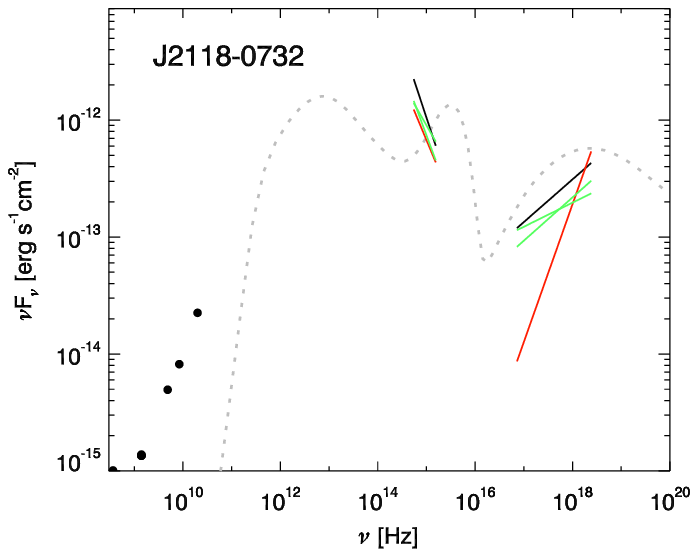}
	\end{tabular}
	\caption{%
	SEDs of the $\gamma$-ray NLS1 galaxies in our sample,  
	built using simultaneous Swift UVOT and XRT data in the observer's frame. Simultaneous UVOT/XRT data are marked in the same colour.   
	The UVOT data are corrected for Galactic extinction. 
	The X-ray spectra are plotted using the best-fit model parameters. 
	The black points represent radio data collected from NED, 
	including measurements by 
	\citet{1986ApJS...61....1B}, 
	\citet{1990PKS...C......0W}, 
	\citet{1991ApJS...75....1B}, 
	\citet{1991ApJS...75.1011G}, 
	\citet{1992ApJS...79..331W}, 
	\citet{1997A&AS..122..235L}, 
	\citet{1998AJ....115.1693C}, 
	\citet{2007ApJS..171...61H}, 
	\citet{2011ApJS..194...29R} 
	and 
	\citet{2014MNRAS.438.3058R}. 
	The gray dotted line represents the SED model of PMN~J0948+0022 taken from \citet{2012A&A...548A.106F} and shifted vertically for comparison. In the case of 1H~0323+342 the large number of available UVOT spectra overlap strongly in this plot. 
	}
\label{fig:swsed}
\end{figure*}

For visually inspecting the correlation between the evolution of optical/UV and X-ray spectra, 
we build the spectral energy distributions for these sources with, if any, 
the UVOT data and the best-fit model of the simultaneous X-ray spectra in the same colors (Figure~\ref{fig:swsed}). 
Data are represented as simple powerlaw connections between the respective UVOT fluxes:
For 1H~0323+342, PMN~J0948+0022 and SDSS~J2118$-$0732, 
the extinction-corrected $v$ and $w2$ band data are used. 
For SDSS~J1222+0413, the $b$ band is used instead of $v$ as there is only two $v$-band exposures. 
For PKS~1502+036 and PKS~2004-447, the $b$ and $u$ band data are used, respectively, 
as the data points in other optical band are too few. 
For SDSS~J0946+1017, the $w1$ and $w2$ band data are used as in the optical band this source is only detected in the $u$ band in one of the observations by \swift.

As can be seen in Figure~\ref{fig:swsed}, 
the broad-band spectral shape varies along with the flux variability. 
The optical/UV and X-ray flux of 1H~0323+342 reveal a simultaneous increase or decrease behavior. 
Its optical/UV emission is becoming bluer when brighter 
indicating the brightening of the source in the optical/UV
being mainly due to the emergence and increase of the emission component in the ultraviolet. 
In comparison, the optical/UV emission of PMN~J0948+0022 shows a redder when brighter trend 
indicating that the increase of the flux is more dominated by the emission in the infrared band. 
For PKS~1502+036, 
the relation of the optical/UV and X-ray spectral variability is not obvious by visual inspection. 
The X-ray spectra vary from steep to flat,
but without strong constraints given 
the low photon statistics. 
For PKS~2004-447, 
the optical/UV and X-ray fluxes show a coordinated increasing/decreasing behavior, 
and the main contributing emission to the flux variability is likely in the infrared and in the X-ray/$\gamma$-ray bands. 
It is hard to see any trends in the other three sources 
as they have very few simultaneous SEDs. 
Both SDSS~J0946+1017 and SDSS~J2118$-$0732 have very red optical/UV spectra, 
indicating a peak at infrared wavelengths. 
SDSS~1222+0413 shows strong optical/UV spectral variability, 
between a bluer spectrum in the bright state and a redder one in the fainter state, 
while the X-ray spectra were hard.

\subsection{Stacked X-ray spectra}
\label{sec:stacking}

The spectra are stacked as 
the statistics of each individual XRT spectrum is low. 
But simply stacking all the spectra may produce incorrect or misleading results in those cases where the sources go through multiple different spectral states.  
As can be seen, 
the sources have shown obvious large amplitude flux variability (Figure~\ref{fig:lc_1h0323}-\ref{fig:lc_j2118}) and significant spectral variability (Figure~\ref{fig:swsed}). 
The spectral indices are different within the uncertainties between  individual observations. 
For example, 
although the largest and smallest $\Gamma$ of PMJ~J0948+0022 have large errors, 
the tightest constraints are $\Gamma=1.2^{+0.2}_{-0.3}$ and $\Gamma=2.2^{+0.3}_{-0.2}$, implying significant spectral variability within the measurement errors.

For 1H~0323+342 and PMN~J0948+0022,
there is an obvious correlation between the optical/UV and X-ray emission, 
and the evolution of broad-band spectral shape \aox\ and the X-ray flux.
For PKS~1502+036, the correlation is not visually obvious from the light curve. 
There is still an trend between the \aox\ and the 2--10\,keV flux, but this trend disappears for the soft X-rays. 
While for PKS~2004-447 the correlation is not visually obvious either for the optical/UV and X-ray flux, or for the \aox\ and X-ray flux. 
So we firstly perform the stacking of the X-ray spectra for 1H~0323+342 and PMN~J0948+0022. 
For 1H~0323+342, as shown in literature, 
the X-ray flux below 10\,keV is usually dominated by emission from the accretion disc and the jet only starts to dominate at $\gtrsim$10\,keV \citep{2015AJ....150...23Y}. 
For PMN~J0948+0022, 
the accretion disc emission seems to only detectable in the soft X-ray band 
and that the jet likely dominates the hard X-ray band according to the decomposition of its broadband X-ray spectrum \citep{2019A&A...632A.120B}. 
Thus, the spectra for stacking are selected according to the soft X-ray fluxes \fsoft\ (0.3--2\,keV).
The spectra with highest and lowest \fsoft\ are selected for generating the stacked spectra at high and low flux states, respectively. 
In addition, 
considering that both the accretion disc and jet may contribute to the X-ray flux, 
the dominant contributor of the X-ray emission is possibly different at different epochs even though they are in similar overall flux state, 
but the main contributing mechanism is unlikely to change in a short time or a stationary phase. 
Therefore, we also stack the spectra within a continuous period of time when \fsoft\ is near a peak (T1) as the high flux state and when \fsoft\ is in a valley (T2) as the low flux state (shaded areas in Figure~\ref{fig:lc_1h0323} and \ref{fig:lc_0948}). 

To compare the spectral properties of high and low flux state, 
the number of spectra for stacking is chosen such that the stacked spectra have roughly equal statistics. 
The more spectra are selected for stacking in each state, 
the higher statistics do the stacked spectra have. 
On the other hand, 
the more spectra are selected 
we will more possibly get misleading results as we may stack the spectra with different spectral shapes. 
We select the spectra for stacking so that 
the stacked spectra have photon counts of $\sim$3000 for 1H~0323+342, 
and $\sim$900 for PMN~J0948+0022. 

According to the SED decomposition of these sources in literature, 
their optical/UV emission is likely dominated by the accretion disc most of the time, 
while the X-rays are more complicated 
and the decomposition is model dependent. 
Study of the spectral evolution when the relative fraction of X-ray emission increases (flatter UV/X-ray slope, i.e. smaller \aox) or decreases (steeper UV/X-ray slope, i.e. larger \aox)
is important to explore whether the accretion disc or the jet makes more contribution to the observed X-rays at different states. 
Thus, 
the spectra for stacking
are also selected based on the broad band spectral slope \aox: 
the X-ray spectra from epochs with largest \aox\ and smallest \aox\ are stacked, respectively. 
Again, the stacked spectra with large \aox\ and small \aox\ have roughly equal statistics as mentioned above.

\begin{figure}
  \centering
  \includegraphics[width=0.95\columnwidth]{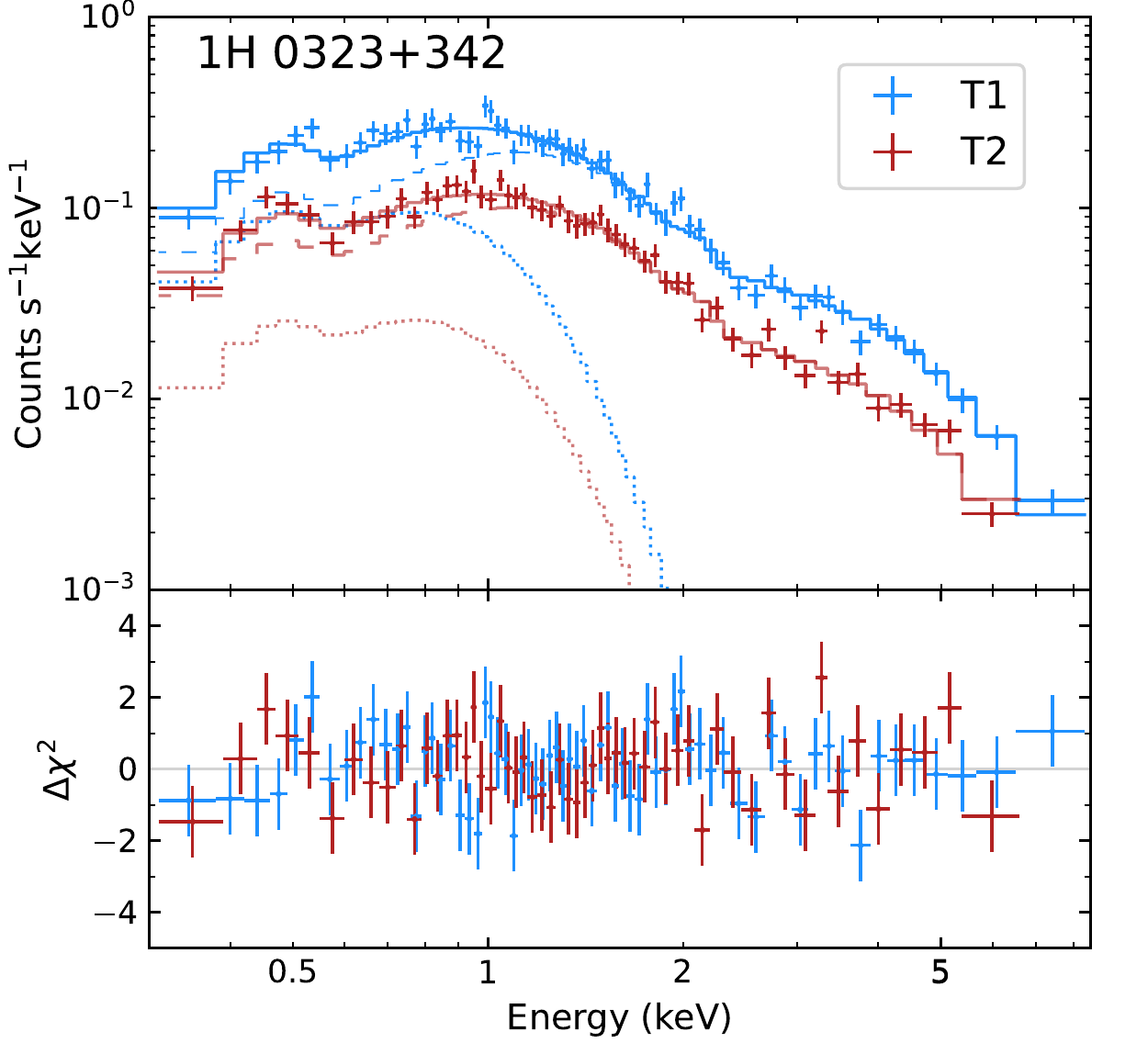}
  \caption{
  Stacked spectra and their best-fit models (upper panel) of 1H~0323+342 at ``T1'' (blue) and ``T2'' (red).
  The best-fit model, the power-law component and the blackbody component, are represented by solid, dashed and dotted lines, respectively. 
  The residuals of the model fit to the spectra are shown in the lower panel.
  }
  \label{fig:1h0323spec}
\end{figure}

The spectra are stacked using the task \textit{addspec}, 
which automatically generated the corresponding response files. 
Then the spectra are grouped so that they have at least 25 counts in each bin. 
A blackbody + power-law model is adopted to fit all the stacked spectra of 1H~0323+342, 
while a single power-law model is adequate to well fit all the stacked spectra of PMN~J0948+0022. 
The Galactic neutral hydrogen absorption is included during the fitting. 
The best-fit parameters are summarized for 1H0323+342 and PMN~J0948+0022 in Table~\ref{tab:stacked1h0323} and \ref{tab:stackedj0948}, respectively. 
As an example we show the stacked spectra and the best-fit models for 1H~0323+342 at peak (T1) and valley (T2) in Figure~\ref{fig:1h0323spec}.

For PKS~1502+036, 
the X-ray light curve does not show any visually obvious `peak' or `valley'. 
So we do not choose such continuous periods as the high or low flux state. 
The spectra for stacking are selected only based on \fsoft\ and \aox, 
and each stacked spectrum has photon counts of $\sim200$. 
For PKS~2004-447, 
there is neither a visually obvious `peak'/`valley' period nor a correlation between \aox\ and the X-ray flux. 
The spectra are only selected based on \fsoft\ and each stacked spectrum has photon counts of $\sim300$. 
A single power-law model with a Galactic neutral hydrogen absorption is adopted to fit the stacked spectra of PKS~1502+036 and PKS~2004-447, 
and the best-fit parameters are summarized in Table~\ref{tab:stacked1502} and \ref{tab:stacked2004}, respectively. 
For the other three sources in the sample, no spectral stacking is performed, since they were observed only a few times. 

\begin{table}
\setlength{\tabcolsep}{3pt}
\footnotesize
\caption{Results of 1H~0323+342 blackbody+power law spectral fit to the stacked spectra}\label{tab:stacked1h0323}
\begin{center}             
\renewcommand{\arraystretch}{1.2}
\begin{tabular}{%
            c
            cccccc
            }
    \hline
    \hline                 
    \multicolumn{1}{c}{State} & 
    \multicolumn{1}{c}{$\Gamma$}  & 
    \multicolumn{1}{c}{$kT$}  &
    \multicolumn{1}{c}{$F_{\rm bb}$}  &
    \multicolumn{1}{c}{$F_{\rm pl}$}  &
    \multicolumn{1}{c}{$R$} & 
    \multicolumn{1}{c}{$\chi^{2}$/dof}
    \\
    \hline
    T1 & 1.74$^{+0.06}_{-0.06}$ & 160$^{+9}_{-9}$ & 43.9$^{+3.2}_{-3.2}$ & 98.2$^{+4.2}_{-4.1}$ & 44.8\% & 58.8/64 \\
    T2 & 1.85$^{+0.07}_{-0.07}$ & 162$^{+17}_{-18}$ & 9.8$^{+1.3}_{-1.3}$ & 43.3$^{+1.8}_{-1.7}$ & 22.7\% & 50.3/52 \\
    High flux & 1.89$^{+0.07}_{-0.07}$ & 150$^{+13}_{-14}$ & 52.3$^{+6.1}_{-6.0}$ & 182.3$^{+8.3}_{-8.0}$ & 28.7\% & 54.3/53  \\
    Low flux & 1.75$^{+0.06}_{-0.06}$ & 170$^{+40}_{-38}$  & 4.0$^{+1.1}_{-1.1}$ & 41.5$^{+1.4}_{-1.3}$ & 9.7\%  & 75.7/60\\
    Flat \aox & 1.88$^{+0.06}_{-0.06}$ & 151$^{+13}_{-14}$ & 47.8$^{+5.6}_{-5.6}$ & 179.6$^{+7.5}_{-7.3}$ & 26.6\%  & 44.2/57 \\
    Steep \aox & 1.68$^{+0.06}_{-0.06}$ & 171$^{+15}_{-16}$ & 9.2$^{+1.1}_{-1.1}$ & 41.3$^{+1.4}_{-1.4}$ & 22.3\% & 66.5/62  \\
    \hline 
\end{tabular}
\parbox[]{\columnwidth}{%
    {\it Notes.}
    A Galactic absorption of column density of $N^{\rm Gal}_{\rm H}=1.17\times10^{21}\rm\,cm^{-2}$ is always included during the fitting. 
    Column 2: Photon index. 
    Column 3: Black body temperature in units of eV. 
    Column 4: Flux of the black body component in the band 0.3--2\,keV in units of $10^{-13}$\,\flux. 
    Column 5: Flux of power law component in the band 0.3--2\,keV in units of $10^{-13}$\,\flux. 
    Column 6: Flux ratio of black body to power law in the band 0.3--2\,keV. 
    Column 7: $\chi^{2}$ and the degrees of freedom. 
    }
\end{center}
\end{table}

\begin{table}
\setlength{\tabcolsep}{4pt}
\footnotesize
\caption{Results of PMN~J0948+0022 power law spectral fits}\label{tab:stackedj0948}
\begin{center}             
\renewcommand{\arraystretch}{1.2}
\begin{tabular}{%
            c
            cccccc
            }
    \hline
    \hline                 
    \multicolumn{1}{>{\centering}m{0.3\columnwidth}}{State} & 
    \multicolumn{1}{>{\centering}m{0.3\columnwidth}}{$\Gamma$}  & 
    \multicolumn{1}{>{\centering}m{0.3\columnwidth}}{$\chi^{2}$/dof}
    \\
    \hline
    T1 & 1.54$^{+0.05}_{-0.05}$ & 30.4/36 \\
    T2 & 1.74$^{+0.05}_{-0.05}$ & 33.5/35 \\
    High flux & 1.58$^{+0.05}_{-0.05}$ & 31.0/34  \\
    Low flux & 1.69$^{+0.05}_{-0.05}$ & 24.0/33 \\
    Flat \aox & 1.63$^{+0.04}_{-0.04}$ & 41.5/47 \\
    Steep \aox & 1.82$^{+0.05}_{-0.05}$ & 41.8/34  \\
    \hline 
\end{tabular}
\parbox[]{\columnwidth}{%
    {\it Notes.}
    A Galactic absorption of column density of $N^{\rm Gal}_{\rm H}=4.73\times10^{20}\rm\,cm^{-2}$ is always included during the fitting. 
    Column 2: Photon index. 
    Column 3: $\chi^{2}$ and the degrees of freedom. 
    }
\end{center}
\end{table}

\begin{table}
\setlength{\tabcolsep}{4pt}
\footnotesize
\caption{Results of PKS~1502+036 spectral fits}\label{tab:stacked1502}
\begin{center}             
\renewcommand{\arraystretch}{1.2}
\begin{tabular}{%
            c
            cccccc
            }
    \hline
    \hline
    \multicolumn{1}{>{\centering}m{0.3\columnwidth}}{State} & 
    \multicolumn{1}{>{\centering}m{0.3\columnwidth}}{$\Gamma$}  & 
    \multicolumn{1}{>{\centering}m{0.3\columnwidth}}{$\chi^{2}$/dof}
    \\
    \hline
    High flux  & 2.06$^{+0.13}_{-0.13}$ & 4.9/7 \\
    Low flux  & 1.20$^{+0.12}_{-0.12}$ & 1.1/7 \\
    Flat \aox & 1.51$^{+0.11}_{-0.11}$ & 5.9/7 \\
    Steep \aox & 2.04$^{+0.17}_{-0.17}$ & 16.4/7  \\
    \hline 
\end{tabular}
\parbox[]{\columnwidth}{%
    {\it Notes.}
    A Galactic absorption of column density of $N^{\rm Gal}_{\rm H}=3.47\times10^{20}\rm\,cm^{-2}$ is always included during the fitting. 
    Column 2: Photon index. 
    Column 3: $\chi^{2}$ and the degrees of freedom.
    }
\end{center}
\end{table}

\begin{table}
\setlength{\tabcolsep}{4pt}
\footnotesize
\caption{Results of PKS~2004-447 spectral fits}\label{tab:stacked2004}
\begin{center}             
\renewcommand{\arraystretch}{1.2}
\begin{tabular}{%
            c
            cccccc
            }
    \hline
    \hline
    \multicolumn{1}{>{\centering}m{0.3\columnwidth}}{State} & 
    \multicolumn{1}{>{\centering}m{0.3\columnwidth}}{$\Gamma$}  & 
    \multicolumn{1}{>{\centering}m{0.3\columnwidth}}{$\chi^{2}$/dof}
    \\
    \hline
    High flux  &  1.60$^{+0.09}_{-0.09}$ & 10.2/11 \\
    Low flux  & 1.44$^{+0.09}_{-0.09}$ & 9.5/12 \\
    \hline 
\end{tabular}
\parbox[]{\columnwidth}{%
    {\it Notes.}
    A Galactic absorption of column density of $N^{\rm Gal}_{\rm H}=2.97\times10^{20}\rm\,cm^{-2}$ is always included during the fitting. 
    Column 2: Photon index. 
    Column 3: $\chi^{2}$ and the degrees of freedom.
    }
\end{center}
\end{table}

\subsection{Notes on individual objects}

Here, we provide a short overview of the sources of our sample, and also highlight similarities and differences in their multiwavelength properties, their black hole masses, and redshifts.

\paragraph{1H~0323+342.} 
This  galaxy is one of the early identified $\gamma$-ray NLS1 galaxies and the nearest one. 
1H~0323+342 has a black hole mass of only a few $10^{7}$\,\msun\ estimated by several different methods including the reverberation mapping method \citep[e.g.][]{2007ApJ...658L..13Z, 
2015AJ....150...23Y,
2018ApJ...866...69P,
2016ApJ...824..149W}. 
The optical spectrum of 1H~0323+342 shows strong Fe{\sc\,ii} multiplets, 
and the narrow emission lines are all weak
\citep[e.g.][]{2007ApJ...658L..13Z, 2014ApJ...795...58L, 2017MNRAS.464.2565L, 2016ApJ...824..149W}. 
Its host galaxy image has revealed a Sersic bulge and a possible disc component \citep[][]{2020MNRAS.492.1450O}. 
A one-armed spiral or ring-like structure is present in both near-infrared and HST optical images, 
suggesting a recent galaxy interaction \citep[][]{2007ApJ...658L..13Z, 2008A&A...490..583A, 2014ApJ...795...58L, 2020MNRAS.492.1450O} 
and making 1H~0323+342 an interesting case to study the triggering of AGN activity.
1H~0323+342 is highly variable across the electromagnetic spectrum \citep[e.g.,][]{2015AJ....150...23Y, 2020MNRAS.496.2213D} and shows X-ray spectral complexity
\citep[e.g.][]{2015AJ....150...23Y, 2020MNRAS.496.2922M}. 

In 2019--2021, we detect a broad flare, more pronounced in the UV and with an amplitude of $\sim$ 1 magnitude. 
As can be seen from the individual and stacked X-ray spectra and the broad-band SED in Figure~\ref{fig:swsed} and \ref{fig:1h0323spec}, we find a 
significant excess of soft X-rays over a single powerlaw. The spectrum of 1H~0323+342 is well fitted by a two-component model, where the soft X-ray emission likely represents emission related to an accretion disc, or is otherwise mimicked by the presence of ionized absorption. 

\paragraph{PMN~J0948+0022.} Another prototype of the $\gamma$-ray emitting NLS1 galaxies is PMN~J0948+0022 \citep[]{2009ApJ...699..976A}.
Similar to 1H~0323+342, 
the optical spectrum of PMN~0948+0022 shows strong Fe{\sc\,ii} multiplets and very weak narrow emission lines 
\citep{2003ApJ...584..147Z}. 
By using width of H$\beta$ and the optical luminosity, and the empirical scaling relations,
the black hole mass of PMN~J0948+0022 is estimated to be a few $10^{7}$\,\msun\ \citep[][]{2003ApJ...584..147Z, 2022RAA....22g5001X}. 
The modeling of PMN~J0948+0022's host galaxy image has shown a brightness profile with Sersic index of $n<2$, 
suggesting a disc-like morphology. 
But unlike 1H~0323+342, 
no interaction feature is detected in this source \citep[][]{2020MNRAS.492.1450O}. 
\citet{2019A&A...632A.120B} have reported that the \xmm\ and \nustar\ joint spectra are characterized by a soft excess on top of a hard power law with a hard photon index, 
indicating a combination of emission from the accretion disk and the jet. 
We find, that a single power law (absorbed with Galactic absorption) is adequate to provide a good fit to the \swift\ X-ray spectra discussed here. In short exposures, multicomponent models cannot be distinguished with \swift. However, the majority of X-ray spectral indices we measure for PNM~J0948+0022 are $<2$, with $\Gamma_{\rm X}$ as low as 1.1, indicating that its X-ray spectrum is hard and the X-ray emission is dominated by the jet most of the time. Only in very few cases we find steeper spectral indices up to 2.3, making a disc contribution likely at those spectral states. 
The X-ray flux and the \aox\ of PMN~J0948+0022 have a similar variability trend as seen for 1H0323+342 (upper panel of Figure~\ref{fig:aox}).

\paragraph{PKS~1502+036.} This is the NLS1 galaxy with the lowest reported SMBH mass of the sample discussed here. A black hole mass of $4\times10^{6}$\,\msun\ was obtained when employing single-epoch optical spectroscopy and empirical scaling relations \citep[][]{2008ApJ...685..801Y}. 
Its host galaxy image 
is well fitted by a disc-like galaxy \citep[][]{2020MNRAS.492.1450O}, however, 
an elliptical host was favored by \citet[][]{2018MNRAS.478L..66D} instead, and  a much higher black hole mass of $\sim7\times10^{8}$\,\msun\ was obtained from bulge luminosity in that study.  
There is a companion close to PKS~1502+036 in the $J$ band image but the source does not show any signs of interaction \citep[][]{2020MNRAS.492.1450O}.

The optical/UV is likely dominated by the emission from the accretion disc in most cases but could also by jet during flares when the optical/UV spectrum becomes apparently redder (Figure~\ref{fig:swsed}). 
Compared to other sources in the sample, 
PKS~1502+036 is relatively fainter in X-rays. 
This can be seen from its \aox\ (Figure~\ref{fig:aox}) and its broad-band SED (Figure~\ref{fig:swsed}). 
The best fit to the X-ray spectrum of PKS~1502+036 is provided by a single power law with Galactic absorption either for individual spectra or stacked ones, consistent with the results reported by \citet[][]{2019A&A...632A.120B}, 
and no intrinsic absorption is detected. 
With respect of the broad band spectral variation, 
the \aox\ and the X-ray flux in the band 2--10\,keV still follow a similar trend as in 1H~0323+342 and PMN~J0948+0022, 
but this trend is not seen for X-rays below 2\,keV.
More simultaneous optical/UV and X-ray monitoring with higher S/N on PKS~1502+036 is needed to explore the broad-band spectral evolution. 

\paragraph{PKS~2004$-$447 and SDSS~J2118$-$0732.}
Among the remaining sources in the sample, 
PKS~2004$-$447 and SDSS~J2118$-$0732 have black hole mass estimates of  a few $10^{7}$\,\msun\ \citep[][]{2021A&A...649A..77G, 2018MNRAS.477.5127Y}. 
Although they were also identified as NLS1 galaxies, 
their spectra are characterized by much stronger Balmer emission lines from the narrow-line region (NLR) compared with other sources in the sample 
(see optical spectrum presented in \citealt{2001ApJ...558..578O} and \citealt{Berton2021} for PKS~2004$-$447, and in \citealt{2018MNRAS.477.5127Y} and \citealt{2020A&A...636L..12J} for SDSS~J2118$=$0732). 
PKS~2004$-$447 is a compact steep-spectrum (CSS) radio source \citep{Gallo2006}.
Its host galaxy appears to be a barred disc galaxy from the best-fit model to its infrared images \citep[][]{2016ApJ...832..157K}, 
while the SDSS image of the host galaxy of SDSS~J2118$-$0732 is well resolved and shows the very characteristic structures of an ongoing galaxy interaction or merger
\citep[][]{2018MNRAS.477.5127Y, 2018rnls.confE..15K}.
Based on the analysis of the SED and/or the X-ray spectra at past epochs, 
the X-rays of these two sources 
are likely dominated by emission from the jet and the disc only has little contribution if not absent \citep[][]{2016A&A...585A..91K, 2018MNRAS.477.5127Y, 2019A&A...632A.120B}.

As shown in our SED of PKS~2004$-$447 and SDSS~J2118$-$0732 (Figure~\ref{fig:swsed}), 
their X-ray spectra are very flat indicating the dominance of jet contribution, consistent with previous findings. 
In particular, the optical/UV spectra are very red, indicating that the emission from accretion disc contributes little to the optical/UV continuum if any. Instead, 
the optical/UV continuum is likely dominated by synchrotron emission from the jet for most of the time. 
There is no obvious trend for the broad-band spectral shape evolution for PKS~2004$-$447 based on its \aox\ vs flux, and the data points are too few to perform this study for SDSS~J2118$-$0732. 
More intense simultaneous multiband monitoring is required to explore the interplay between accretion and jet.

\paragraph{SDSS~J0946+1017 and SDSS~J1222+0413.}
These two NLS1 galaxies 
have spectral features similar to PMN~J0948+0022 or PKS~1502+036, strong FeII and weak narrow emission lines, 
but they have black hole masses of a few $10^{8}$\,\msun\ \citep[][]{2015MNRAS.454L..16Y, 2019MNRAS.487L..40Y}, 
significantly higher than the other sources. 
They are of particular interest, because they represent the high-redshift population of our sample, and are among the highest-redshift 
$\gamma$-ray emitting NLS1 galaxies identified so far \citep[][]{2015MNRAS.454L..16Y, 2019MNRAS.487L..40Y, 2021MNRAS.504L..22R}. 
Their host galaxies are not yet detected due to large distances ($z\approx1$) \citep[][]{2020MNRAS.492.1450O}.  
Based on the analysis of the SED and/or the X-ray spectra at past epochs, 
the X-rays of these two galaxies are likely dominated by emission from the jet and the disc only has little contribution if not absent  
\citep[][]{2015MNRAS.454L..16Y, 2019MNRAS.487L..40Y, 2019A&A...632A.120B}. 
Due to their high redshift, any soft X-ray accretion-disc contribution, even if intrinsically present, would be likely missed since shifted out of the observable band. 

As shown in Figure~\ref{fig:swsed}, 
the \swift/UVOT has caught strong optical/UV spectral variability of SDSS~J1222+0413, 
from a very blue spectrum at high flux state indicating a strong contribution from the accretion disc, to a red spectrum. %
This indicates that the accretion disc has faded strongly. 
Similarly, the UV spectra of SDSS~J0946+1017 are very red during \swift\ observations although there is a lack of simultaneous optical detections.

\section{Discussion} 
\label{sec:discussion}

Only a small number of NLS1 galaxies, about 20, has been detected in the $\gamma$-ray band so far
\citep[see][for a review]{2018rnls.confE..15K}. 
These sources provide us with important new laboratories of the disc-jet connection in a previously unexplored parameter space. 
\swift\ in particular delivers simultaneous SEDs from the optical to the X-rays  \citep[e.g.,][]{Grupe2010, 2015AJ....150...23Y, 2020MNRAS.496.2213D}.  
Since all $\gamma$-ray NLS1 galaxies are highly variable across the electromagnetic spectrum, important questions are if their SEDs show the same spectral components at all times, or if there is systematic trends, either with brightness state, or with time, and
also if there are significant dependencies on black hole mass or on accretion rate 
$\dot{m}/\dot{m}_{\rm Edd}$ where $\dot{m}_{\rm Edd}$ is the Eddington accretion rate. 
An important question in this respect regards the contribution of thermal (accretion disc) versus non-thermal (jet-related) processes to their SEDs. %

Of particular interest is the identification of NLS1 galaxies, that show both, accretion disc and jet components in their SEDs. 
Because NLS1 galaxies have, on average, lower SMBH masses and therefore faster accretion-disc variability timescales than their classical blazar counterparts, they are particularly well suited in studying the jet-disc symbiosis across short timescales  \citep{2018rnls.confE..15K}.   
We have therefore analyzed all available \swift\ data of recent years for a small sample of $\gamma$-ray emitting NLS1 galaxies, and have also added their long-term \swift\ light curve evolution, and used these multiyear data for various spectral analyses. %

Most of the NLS1 galaxies of our study have shown significant variability in optical/UV and X-rays, including events of rapid flaring. An exception is
J0946+1017 which may be due to the sparsity of observations. 
In addition to the flux variability, there is also strong UV--optical and X-ray spectral variability in most of the sources. 
X-ray spectral index measurements reveal exceptionally flat X-ray spectra for most of the galaxies most of the time, in marked contrast to radio-quiet NLS1 galaxies that are characterized by ultra-steep soft X-ray spectra \citep[e.g.,][]{Puchnarewicz1992, Grupe1998}, interpreted as signs of reprocessing from an accretion disc and/or ionized absorption of the intrinsic emission \citep[see][for a review]{Komossa2008}. The exceptionally flat X-ray spectra therefore already indicate a strong contribution of non-thermal jet emission in the X-ray band in the majority of sources. %

The different patterns of the soft and hard X-ray flux variability of the NLS1 galaxies studied here, and their SEDs at different epochs, are evident in Figure~\ref{fig:swsed}. 
However, 
there is no homogeneous trend throughout the sample. 
The optical/UV spectra become bluer when the flux increases in 1H~0323+342 and SDSS~J1222+0413. 
The X-ray spectral slope of 1H~0323+342 becomes slightly steeper when flux increases. 
Considering that the optical/UV emission of these two sources is dominated by the contribution from a strong accretion disc component based on the modeling and decomposition of their broad band SEDs \cite[e.g.][]{2009ApJ...707L.142A, 2015MNRAS.454L..16Y, 2019MNRAS.487L..40Y}, 
the spectral variability behavior in these two sources is likely due to the variable emission from the accretion disc. 

For PMN~J0948+0022 and PKS~1502+036, 
the decomposition of their SEDs has revealed the dominance of strong thermal emission in the optical/UV from the accretion disc by previous work, 
whereas non-thermal emission from the jet component likely dominates their X-ray spectra
\citep[][]{2009ApJ...707L.142A, 2012A&A...548A.106F, 2016ApJ...820...52P}. 
At peak luminosities the optical/UV spectra of PMN~J0948+0022 and PKS~1502+036 become obviously redder, 
suggesting that the brightening is mainly due to the increasing contribution from emission peaked in the radio/infrared band, i.e., synchrotron emission from the jet
\citep[e.g.][]{2015MNRAS.446.2456D, 2016MNRAS.463.4469D, 2016ApJ...820...52P}. 
This is in stark contrast to the variability behavior reported above of 1H~0323+342 and SDSS~J1222+0413.  

At selected epochs, a decomposition of the broad band SED of PKS~2004$-$447  \citep[e.g.][]{2009ApJ...707L.142A, 2021A&A...649A..77G} and SDSS~J2118$-$0732 \citep{2018MNRAS.477.5127Y}
has shown  
that the contribution from an accretion disc is very weak at those epochs.
Here, we find that  PKS~2004$-$447 and SDSS~J2118$-$0732 have red optical/UV spectra and very hard X-ray spectra, 
indicating that the emission peaks at infrared and $\gamma$-rays, respectively, 
and that the emission from the jet dominates at all observed frequencies.

\section{Summary and conclusions}
\label{sec:summary}

In this work we have studied the \swift\ monitoring data up to 2021 December of seven bright $\gamma$-ray detected NLS1 galaxies. 
The simultaneous optical/UV and X-ray observations by \swift\ provide good data sets to investigate the variability of the emission from the coupled accretion and jet process. 
Among the sample, the accretion disc dominates the emission in the optical/UV and makes significant contributions to the X-rays of 1H~0323+342 as derived from its X-ray spectra and broad-band SEDs. The jet 
contribution may only be significant in hard X-rays. 
In other sources, the optical/UV emission is mainly from the accretion disc but could also be dominated by synchrotron emission during flares,  
while their X-ray spectra are always dominated by the emission from the jet and the accretion disc makes very little contribution even though some of the galaxies have similar black hole masses and host galaxies and some do not. 
A decreasing trend of \aox\ with increasing X-ray flux can be seen in 1H~0323+342, PMN~J0948+0022 and PKS~1502+036 although the main contributors to X-rays in these source are different. 
However, this trend is not seen in PKS~2004-447 (Figure~\ref{fig:aox}). 
There is no adequate number of simultaneous UV and X-ray observations for the other three sources to explore this trend. 
More simultaneous multifrequency monitoring observations of these $\gamma$-ray detected NLS1 galaxies are needed to investigate their optical/UV and X-ray emission variability in order to study the accretion disk/jet evolution in this rare species.

\section*{Acknowledgements}

We thank our anonymous referee for very useful comments.
SY acknowledges the support by an Alexander von Humboldt Foundation Fellowship. 
This research has made use of data products from \swift\ and software provided by the High Energy Astrophysics Science Archive Research Center (HEASARC), which is a service of the Astrophysics Science Division at NASA/GSFC and the High Energy Astrophysics Division of the Smithsonian Astrophysical Observatory. 
This research also made use of XRT Data Analysis Software (XRTDAS) developed under the responsibility of the AOS Science Data Center (ASDC), Italy. 
This work has also used the NASA Astrophysics Data System Abstract Service (ADS), and the NASA/IPAC Extragalactic Database (NED) which is operated by the Jet Propulsion Laboratory, California Institute of Technology, under contract with the National Aeronautics and Space Administration.


\section*{Data Availability}

The data in this paper are available in 
the astronomical archives of HEASARC at
\url{https://heasarc.gsfc.nasa.gov/docs/archive.html}
and the \swift\ archive at \url{https://swift.gsfc.nasa.gov/archive/}, 
and can be accessed with the source coordinates or observation IDs.




\bibliographystyle{mnras}
\bibliography{reference} 

\begin{thebibliography}{}
\makeatletter
\relax
\def\mn@urlcharsother{\let\do\@makeother \do\$\do\&\do\#\do\^\do\_\do\%\do\~}
\def\mn@doi{\begingroup\mn@urlcharsother \@ifnextchar [ {\mn@doi@}
  {\mn@doi@[]}}
\def\mn@doi@[#1]#2{\def\@tempa{#1}\ifx\@tempa\@empty \href
  {http://dx.doi.org/#2} {doi:#2}\else \href {http://dx.doi.org/#2} {#1}\fi
  \endgroup}
\def\mn@eprint#1#2{\mn@eprint@#1:#2::\@nil}
\def\mn@eprint@arXiv#1{\href {http://arxiv.org/abs/#1} {{\tt arXiv:#1}}}
\def\mn@eprint@dblp#1{\href {http://dblp.uni-trier.de/rec/bibtex/#1.xml}
  {dblp:#1}}
\def\mn@eprint@#1:#2:#3:#4\@nil{\def\@tempa {#1}\def\@tempb {#2}\def\@tempc
  {#3}\ifx \@tempc \@empty \let \@tempc \@tempb \let \@tempb \@tempa \fi \ifx
  \@tempb \@empty \def\@tempb {arXiv}\fi \@ifundefined
  {mn@eprint@\@tempb}{\@tempb:\@tempc}{\expandafter \expandafter \csname
  mn@eprint@\@tempb\endcsname \expandafter{\@tempc}}}

\bibitem[\protect\citeauthoryear{{Abdo} et~al.,}{{Abdo}
  et~al.}{2009a}]{2009ApJ...699..976A}
{Abdo} A.~A.,  et~al., 2009a, \mn@doi [\apj] {10.1088/0004-637X/699/2/976},
  \href {https://ui.adsabs.harvard.edu/abs/2009ApJ...699..976A} {699, 976}

\bibitem[\protect\citeauthoryear{{Abdo} et~al.,}{{Abdo}
  et~al.}{2009b}]{2009ApJ...707L.142A}
{Abdo} A.~A.,  et~al., 2009b, \mn@doi [\apjl] {10.1088/0004-637X/707/2/L142},
  \href {https://ui.adsabs.harvard.edu/abs/2009ApJ...707L.142A} {707, L142}

\bibitem[\protect\citeauthoryear{{Abdo} et~al.,}{{Abdo}
  et~al.}{2010}]{2010ApJS..188..405A}
{Abdo} A.~A.,  et~al., 2010, \mn@doi [\apjs] {10.1088/0067-0049/188/2/405},
  \href {https://ui.adsabs.harvard.edu/abs/2010ApJS..188..405A} {188, 405}

\bibitem[\protect\citeauthoryear{{Angelakis}, {Kiehlmann}, {Myserlis},
  {Blinov}, {Eggen}, {Itoh}, {Marchili}  \& {Zensus}}{{Angelakis}
  et~al.}{2018}]{2018A&A...618A..92A}
{Angelakis} E.,  {Kiehlmann} S.,  {Myserlis} I.,  {Blinov} D.,  {Eggen} J.,
  {Itoh} R.,  {Marchili} N.,   {Zensus} J.~A.,  2018, \mn@doi [\aap]
  {10.1051/0004-6361/201832890}, \href
  {https://ui.adsabs.harvard.edu/abs/2018A&A...618A..92A} {618, A92}

\bibitem[\protect\citeauthoryear{{Ant{\'o}n}, {Browne}  \&
  {March{\~a}}}{{Ant{\'o}n} et~al.}{2008}]{2008A&A...490..583A}
{Ant{\'o}n} S.,  {Browne} I.~W.~A.,   {March{\~a}} M.~J.,  2008, \mn@doi [\aap]
  {10.1051/0004-6361:20078926}, \href
  {https://ui.adsabs.harvard.edu/abs/2008A&A...490..583A} {490, 583}

\bibitem[\protect\citeauthoryear{{Arnaud}}{{Arnaud}}{1996}]{1996ASPC..101...17A}
{Arnaud} K.~A.,  1996, in {Jacoby} G.~H.,  {Barnes} J.,  eds,  Astronomical
  Society of the Pacific Conference Series Vol. 101, Astronomical Data Analysis
  Software and Systems V. p.~17

\bibitem[\protect\citeauthoryear{{Bahcall}, {Kirhakos}, {Saxe}  \&
  {Schneider}}{{Bahcall} et~al.}{1997}]{1997ApJ...479..642B}
{Bahcall} J.~N.,  {Kirhakos} S.,  {Saxe} D.~H.,   {Schneider} D.~P.,  1997,
  \mn@doi [\apj] {10.1086/303926}, \href
  {https://ui.adsabs.harvard.edu/abs/1997ApJ...479..642B} {479, 642}

\bibitem[\protect\citeauthoryear{{Becker}, {White}  \& {Edwards}}{{Becker}
  et~al.}{1991}]{1991ApJS...75....1B}
{Becker} R.~H.,  {White} R.~L.,   {Edwards} A.~L.,  1991, \mn@doi [\apjs]
  {10.1086/191529}, \href
  {https://ui.adsabs.harvard.edu/abs/1991ApJS...75....1B} {75, 1}

\bibitem[\protect\citeauthoryear{{Bennett}, {Lawrence}, {Burke}, {Hewitt}  \&
  {Mahoney}}{{Bennett} et~al.}{1986}]{1986ApJS...61....1B}
{Bennett} C.~L.,  {Lawrence} C.~R.,  {Burke} B.~F.,  {Hewitt} J.~N.,
  {Mahoney} J.,  1986, \mn@doi [\apjs] {10.1086/191108}, \href
  {https://ui.adsabs.harvard.edu/abs/1986ApJS...61....1B} {61, 1}

\bibitem[\protect\citeauthoryear{{Berton}, {Braito}, {Mathur}, {Foschini},
  {Piconcelli}, {Chen}  \& {Pogge}}{{Berton}
  et~al.}{2019}]{2019A&A...632A.120B}
{Berton} M.,  {Braito} V.,  {Mathur} S.,  {Foschini} L.,  {Piconcelli} E.,
  {Chen} S.,   {Pogge} R.~W.,  2019, \mn@doi [\aap]
  {10.1051/0004-6361/201935929}, \href
  {https://ui.adsabs.harvard.edu/abs/2019A&A...632A.120B} {632, A120}

\bibitem[\protect\citeauthoryear{{Berton} et~al.,}{{Berton}
  et~al.}{2021a}]{2021A&A...654A.125B}
{Berton} M.,  et~al., 2021a, \mn@doi [\aap] {10.1051/0004-6361/202141409},
  \href {https://ui.adsabs.harvard.edu/abs/2021A&A...654A.125B} {654, A125}

\bibitem[\protect\citeauthoryear{{Berton} et~al.,}{{Berton}
  et~al.}{2021b}]{Berton2021}
{Berton} M.,  et~al., 2021b, \mn@doi [\aap] {10.1051/0004-6361/202141409},
  \href {https://ui.adsabs.harvard.edu/abs/2021A&A...654A.125B} {654, A125}

\bibitem[\protect\citeauthoryear{{Blandford}, {Meier}  \&
  {Readhead}}{{Blandford} et~al.}{2019}]{Blandford2019}
{Blandford} R.,  {Meier} D.,   {Readhead} A.,  2019, \mn@doi [\araa]
  {10.1146/annurev-astro-081817-051948}, \href
  {https://ui.adsabs.harvard.edu/abs/2019ARA&A..57..467B} {57, 467}

\bibitem[\protect\citeauthoryear{{Boroson}}{{Boroson}}{2002}]{2002ApJ...565...78B}
{Boroson} T.~A.,  2002, \mn@doi [\apj] {10.1086/324486}, \href
  {https://ui.adsabs.harvard.edu/abs/2002ApJ...565...78B} {565, 78}

\bibitem[\protect\citeauthoryear{{Breeveld} et~al.,}{{Breeveld}
  et~al.}{2010}]{2010MNRAS.406.1687B}
{Breeveld} A.~A.,  et~al., 2010, \mn@doi [\mnras]
  {10.1111/j.1365-2966.2010.16832.x}, \href
  {https://ui.adsabs.harvard.edu/abs/2010MNRAS.406.1687B} {406, 1687}

\bibitem[\protect\citeauthoryear{{Burrows} et~al.,}{{Burrows}
  et~al.}{2005}]{2005SSRv..120..165B}
{Burrows} D.~N.,  et~al., 2005, \mn@doi [\ssr] {10.1007/s11214-005-5097-2},
  \href {https://ui.adsabs.harvard.edu/abs/2005SSRv..120..165B} {120, 165}

\bibitem[\protect\citeauthoryear{{Calderone}, {Foschini}, {Ghisellini},
  {Colpi}, {Maraschi}, {Tavecchio}, {Decarli}  \& {Tagliaferri}}{{Calderone}
  et~al.}{2011}]{2011MNRAS.413.2365C}
{Calderone} G.,  {Foschini} L.,  {Ghisellini} G.,  {Colpi} M.,  {Maraschi} L.,
  {Tavecchio} F.,  {Decarli} R.,   {Tagliaferri} G.,  2011, \mn@doi [\mnras]
  {10.1111/j.1365-2966.2011.18308.x}, \href
  {https://ui.adsabs.harvard.edu/abs/2011MNRAS.413.2365C} {413, 2365}

\bibitem[\protect\citeauthoryear{{Condon}, {Cotton}, {Greisen}, {Yin},
  {Perley}, {Taylor}  \& {Broderick}}{{Condon}
  et~al.}{1998}]{1998AJ....115.1693C}
{Condon} J.~J.,  {Cotton} W.~D.,  {Greisen} E.~W.,  {Yin} Q.~F.,  {Perley}
  R.~A.,  {Taylor} G.~B.,   {Broderick} J.~J.,  1998, \mn@doi [\aj]
  {10.1086/300337}, \href
  {https://ui.adsabs.harvard.edu/abs/1998AJ....115.1693C} {115, 1693}

\bibitem[\protect\citeauthoryear{{D'Ammando}}{{D'Ammando}}{2020a}]{2020MNRAS.496.2213D}
{D'Ammando} F.,  2020a, \mn@doi [\mnras] {10.1093/mnras/staa1580}, \href
  {https://ui.adsabs.harvard.edu/abs/2020MNRAS.496.2213D} {496, 2213}

\bibitem[\protect\citeauthoryear{{D'Ammando}}{{D'Ammando}}{2020b}]{2020MNRAS.498..859D}
{D'Ammando} F.,  2020b, \mn@doi [\mnras] {10.1093/mnras/staa2471}, \href
  {https://ui.adsabs.harvard.edu/abs/2020MNRAS.498..859D} {498, 859}

\bibitem[\protect\citeauthoryear{{D'Ammando} et~al.,}{{D'Ammando}
  et~al.}{2015}]{2015MNRAS.446.2456D}
{D'Ammando} F.,  et~al., 2015, \mn@doi [\mnras] {10.1093/mnras/stu2251}, \href
  {https://ui.adsabs.harvard.edu/abs/2015MNRAS.446.2456D} {446, 2456}

\bibitem[\protect\citeauthoryear{{D'Ammando} et~al.,}{{D'Ammando}
  et~al.}{2016}]{2016MNRAS.463.4469D}
{D'Ammando} F.,  et~al., 2016, \mn@doi [\mnras] {10.1093/mnras/stw2325}, \href
  {https://ui.adsabs.harvard.edu/abs/2016MNRAS.463.4469D} {463, 4469}

\bibitem[\protect\citeauthoryear{{D'Ammando}, {Acosta-Pulido}, {Capetti},
  {Baldi}, {Orienti}, {Raiteri}  \& {Ramos Almeida}}{{D'Ammando}
  et~al.}{2018}]{2018MNRAS.478L..66D}
{D'Ammando} F.,  {Acosta-Pulido} J.~A.,  {Capetti} A.,  {Baldi} R.~D.,
  {Orienti} M.,  {Raiteri} C.~M.,   {Ramos Almeida} C.,  2018, \mn@doi [\mnras]
  {10.1093/mnrasl/sly072}, \href
  {https://ui.adsabs.harvard.edu/abs/2018MNRAS.478L..66D} {478, L66}

\bibitem[\protect\citeauthoryear{{Di Matteo}, {Springel}  \& {Hernquist}}{{Di
  Matteo} et~al.}{2005}]{DiMatteo2005}
{Di Matteo} T.,  {Springel} V.,   {Hernquist} L.,  2005, \mn@doi [\nat]
  {10.1038/nature03335}, \href
  {https://ui.adsabs.harvard.edu/abs/2005Natur.433..604D} {433, 604}

\bibitem[\protect\citeauthoryear{{Dunlop}, {McLure}, {Kukula}, {Baum}, {O'Dea}
  \& {Hughes}}{{Dunlop} et~al.}{2003}]{2003MNRAS.340.1095D}
{Dunlop} J.~S.,  {McLure} R.~J.,  {Kukula} M.~J.,  {Baum} S.~A.,  {O'Dea}
  C.~P.,   {Hughes} D.~H.,  2003, \mn@doi [\mnras]
  {10.1046/j.1365-8711.2003.06333.x}, \href
  {https://ui.adsabs.harvard.edu/abs/2003MNRAS.340.1095D} {340, 1095}

\bibitem[\protect\citeauthoryear{{Eggen}, {Miller}  \& {Maune}}{{Eggen}
  et~al.}{2013}]{2013ApJ...773...85E}
{Eggen} J.~R.,  {Miller} H.~R.,   {Maune} J.~D.,  2013, \mn@doi [\apj]
  {10.1088/0004-637X/773/2/85}, \href
  {https://ui.adsabs.harvard.edu/abs/2013ApJ...773...85E} {773, 85}

\bibitem[\protect\citeauthoryear{{Fitzpatrick}}{{Fitzpatrick}}{1999}]{1999PASP..111...63F}
{Fitzpatrick} E.~L.,  1999, \mn@doi [\pasp] {10.1086/316293}, \href
  {https://ui.adsabs.harvard.edu/abs/1999PASP..111...63F} {111, 63}

\bibitem[\protect\citeauthoryear{{Floyd}, {Kukula}, {Dunlop}, {McLure},
  {Miller}, {Percival}, {Baum}  \& {O'Dea}}{{Floyd}
  et~al.}{2004}]{2004MNRAS.355..196F}
{Floyd} D. J.~E.,  {Kukula} M.~J.,  {Dunlop} J.~S.,  {McLure} R.~J.,  {Miller}
  L.,  {Percival} W.~J.,  {Baum} S.~A.,   {O'Dea} C.~P.,  2004, \mn@doi
  [\mnras] {10.1111/j.1365-2966.2004.08315.x}, \href
  {https://ui.adsabs.harvard.edu/abs/2004MNRAS.355..196F} {355, 196}

\bibitem[\protect\citeauthoryear{{Foschini} et~al.,}{{Foschini}
  et~al.}{2012}]{2012A&A...548A.106F}
{Foschini} L.,  et~al., 2012, \mn@doi [\aap] {10.1051/0004-6361/201220225},
  \href {https://ui.adsabs.harvard.edu/abs/2012A&A...548A.106F} {548, A106}

\bibitem[\protect\citeauthoryear{{Foschini} et~al.,}{{Foschini}
  et~al.}{2015}]{Foschini2015}
{Foschini} L.,  et~al., 2015, \mn@doi [\aap] {10.1051/0004-6361/201424972},
  \href {https://ui.adsabs.harvard.edu/abs/2015A&A...575A..13F} {575, A13}

\bibitem[\protect\citeauthoryear{{Gabanyi}, {Moor}  \& {Frey}}{{Gabanyi}
  et~al.}{2018}]{2018rnls.confE..42G}
{Gabanyi} K.,  {Moor} A.,   {Frey} S.,  2018, in Revisiting Narrow-Line Seyfert
  1 Galaxies and their Place in the Universe. p.~42 (\mn@eprint {arXiv}
  {1807.05802})

\bibitem[\protect\citeauthoryear{{Gallo} et~al.,}{{Gallo}
  et~al.}{2006}]{Gallo2006}
{Gallo} L.~C.,  et~al., 2006, \mn@doi [\mnras]
  {10.1111/j.1365-2966.2006.10482.x}, \href
  {https://ui.adsabs.harvard.edu/abs/2006MNRAS.370..245G} {370, 245}

\bibitem[\protect\citeauthoryear{{Gehrels} et~al.,}{{Gehrels}
  et~al.}{2004}]{Gehrels2004}
{Gehrels} N.,  et~al., 2004, \mn@doi [\apj] {10.1086/422091}, \href
  {https://ui.adsabs.harvard.edu/abs/2004ApJ...611.1005G} {611, 1005}

\bibitem[\protect\citeauthoryear{{Gokus} et~al.,}{{Gokus}
  et~al.}{2021}]{2021A&A...649A..77G}
{Gokus} A.,  et~al., 2021, \mn@doi [\aap] {10.1051/0004-6361/202039378}, \href
  {https://ui.adsabs.harvard.edu/abs/2021A&A...649A..77G} {649, A77}

\bibitem[\protect\citeauthoryear{{Gregory} \& {Condon}}{{Gregory} \&
  {Condon}}{1991}]{1991ApJS...75.1011G}
{Gregory} P.~C.,  {Condon} J.~J.,  1991, \mn@doi [\apjs] {10.1086/191559},
  \href {https://ui.adsabs.harvard.edu/abs/1991ApJS...75.1011G} {75, 1011}

\bibitem[\protect\citeauthoryear{{Grupe}, {Beuermann}, {Thomas}, {Mannheim}  \&
  {Fink}}{{Grupe} et~al.}{1998}]{Grupe1998}
{Grupe} D.,  {Beuermann} K.,  {Thomas} H.~C.,  {Mannheim} K.,   {Fink} H.~H.,
  1998, \aap, \href {https://ui.adsabs.harvard.edu/abs/1998A&A...330...25G}
  {330, 25}

\bibitem[\protect\citeauthoryear{{Grupe}, {Komossa}, {Leighly}  \&
  {Page}}{{Grupe} et~al.}{2010}]{Grupe2010}
{Grupe} D.,  {Komossa} S.,  {Leighly} K.~M.,   {Page} K.~L.,  2010, \mn@doi
  [\apjs] {10.1088/0067-0049/187/1/64}, \href
  {https://ui.adsabs.harvard.edu/abs/2010ApJS..187...64G} {187, 64}

\bibitem[\protect\citeauthoryear{{Healey}, {Romani}, {Taylor}, {Sadler},
  {Ricci}, {Murphy}, {Ulvestad}  \& {Winn}}{{Healey}
  et~al.}{2007}]{2007ApJS..171...61H}
{Healey} S.~E.,  {Romani} R.~W.,  {Taylor} G.~B.,  {Sadler} E.~M.,  {Ricci} R.,
   {Murphy} T.,  {Ulvestad} J.~S.,   {Winn} J.~N.,  2007, \mn@doi [\apjs]
  {10.1086/513742}, \href
  {https://ui.adsabs.harvard.edu/abs/2007ApJS..171...61H} {171, 61}

\bibitem[\protect\citeauthoryear{{Hill} et~al.,}{{Hill}
  et~al.}{2004}]{2004SPIE.5165..217H}
{Hill} J.~E.,  et~al., 2004, in {Flanagan} K.~A.,  {Siegmund} O. H.~W.,  eds,
  Society of Photo-Optical Instrumentation Engineers (SPIE) Conference Series
  Vol. 5165, X-Ray and Gamma-Ray Instrumentation for Astronomy XIII. pp
  217--231, \mn@doi{10.1117/12.505728}

\bibitem[\protect\citeauthoryear{{Itoh} et~al.,}{{Itoh}
  et~al.}{2013}]{2013ApJ...775L..26I}
{Itoh} R.,  et~al., 2013, \mn@doi [\apjl] {10.1088/2041-8205/775/1/L26}, \href
  {https://ui.adsabs.harvard.edu/abs/2013ApJ...775L..26I} {775, L26}

\bibitem[\protect\citeauthoryear{{Ivezi{\'c}} et~al.,}{{Ivezi{\'c}}
  et~al.}{2002}]{2002AJ....124.2364I}
{Ivezi{\'c}} {\v{Z}}.,  et~al., 2002, \mn@doi [\aj] {10.1086/344069}, \href
  {https://ui.adsabs.harvard.edu/abs/2002AJ....124.2364I} {124, 2364}

\bibitem[\protect\citeauthoryear{{J{\"a}rvel{\"a}}, {L{\"a}hteenm{\"a}ki}  \&
  {Berton}}{{J{\"a}rvel{\"a}} et~al.}{2018}]{2018A&A...619A..69J}
{J{\"a}rvel{\"a}} E.,  {L{\"a}hteenm{\"a}ki} A.,   {Berton} M.,  2018, \mn@doi
  [\aap] {10.1051/0004-6361/201832876}, \href
  {https://ui.adsabs.harvard.edu/abs/2018A&A...619A..69J} {619, A69}

\bibitem[\protect\citeauthoryear{{J{\"a}rvel{\"a}}, {Berton}, {Ciroi},
  {Congiu}, {L{\"a}hteenm{\"a}ki}  \& {Di Mille}}{{J{\"a}rvel{\"a}}
  et~al.}{2020}]{2020A&A...636L..12J}
{J{\"a}rvel{\"a}} E.,  {Berton} M.,  {Ciroi} S.,  {Congiu} E.,
  {L{\"a}hteenm{\"a}ki} A.,   {Di Mille} F.,  2020, \mn@doi [\aap]
  {10.1051/0004-6361/202037826}, \href
  {https://ui.adsabs.harvard.edu/abs/2020A&A...636L..12J} {636, L12}

\bibitem[\protect\citeauthoryear{{Jiang} et~al.,}{{Jiang}
  et~al.}{2012}]{2012ApJ...759L..31J}
{Jiang} N.,  et~al., 2012, \mn@doi [\apjl] {10.1088/2041-8205/759/2/L31}, \href
  {https://ui.adsabs.harvard.edu/abs/2012ApJ...759L..31J} {759, L31}

\bibitem[\protect\citeauthoryear{{Komossa}}{{Komossa}}{2008}]{Komossa2008}
{Komossa} S.,  2008, in Revista Mexicana de Astronomia y Astrofisica Conference
  Series. pp 86--92 (\mn@eprint {arXiv} {0710.3326})

\bibitem[\protect\citeauthoryear{{Komossa}}{{Komossa}}{2018}]{2018rnls.confE..15K}
{Komossa} S.,  2018, in Revisiting Narrow-Line Seyfert 1 Galaxies and their
  Place in the Universe. p.~15 (\mn@eprint {arXiv} {1807.03666})

\bibitem[\protect\citeauthoryear{{Komossa}, {Voges}, {Xu}, {Mathur}, {Adorf},
  {Lemson}, {Duschl}  \& {Grupe}}{{Komossa} et~al.}{2006}]{2006AJ....132..531K}
{Komossa} S.,  {Voges} W.,  {Xu} D.,  {Mathur} S.,  {Adorf} H.-M.,  {Lemson}
  G.,  {Duschl} W.~J.,   {Grupe} D.,  2006, \mn@doi [\aj] {10.1086/505043},
  \href {https://ui.adsabs.harvard.edu/abs/2006AJ....132..531K} {132, 531}

\bibitem[\protect\citeauthoryear{{Kotilainen}, {Le{\'o}n-Tavares},
  {Olgu{\'\i}n-Iglesias}, {Baes}, {An{\'o}rve}, {Chavushyan}  \&
  {Carrasco}}{{Kotilainen} et~al.}{2016}]{2016ApJ...832..157K}
{Kotilainen} J.~K.,  {Le{\'o}n-Tavares} J.,  {Olgu{\'\i}n-Iglesias} A.,  {Baes}
  M.,  {An{\'o}rve} C.,  {Chavushyan} V.,   {Carrasco} L.,  2016, \mn@doi
  [\apj] {10.3847/0004-637X/832/2/157}, \href
  {https://ui.adsabs.harvard.edu/abs/2016ApJ...832..157K} {832, 157}

\bibitem[\protect\citeauthoryear{{Kreikenbohm} et~al.,}{{Kreikenbohm}
  et~al.}{2016}]{2016A&A...585A..91K}
{Kreikenbohm} A.,  et~al., 2016, \mn@doi [\aap] {10.1051/0004-6361/201424818},
  \href {https://ui.adsabs.harvard.edu/abs/2016A&A...585A..91K} {585, A91}

\bibitem[\protect\citeauthoryear{{Landt} et~al.,}{{Landt}
  et~al.}{2017}]{2017MNRAS.464.2565L}
{Landt} H.,  et~al., 2017, \mn@doi [\mnras] {10.1093/mnras/stw2447}, \href
  {https://ui.adsabs.harvard.edu/abs/2017MNRAS.464.2565L} {464, 2565}

\bibitem[\protect\citeauthoryear{{Laurent-Muehleisen}, {Kollgaard}, {Ryan},
  {Feigelson}, {Brinkmann}  \& {Siebert}}{{Laurent-Muehleisen}
  et~al.}{1997}]{1997A&AS..122..235L}
{Laurent-Muehleisen} S.~A.,  {Kollgaard} R.~I.,  {Ryan} P.~J.,  {Feigelson}
  E.~D.,  {Brinkmann} W.,   {Siebert} J.,  1997, \mn@doi [\aaps]
  {10.1051/aas:1997331}, \href
  {https://ui.adsabs.harvard.edu/abs/1997A&AS..122..235L} {122, 235}

\bibitem[\protect\citeauthoryear{{Le{\'o}n Tavares} et~al.,}{{Le{\'o}n Tavares}
  et~al.}{2014}]{2014ApJ...795...58L}
{Le{\'o}n Tavares} J.,  et~al., 2014, \mn@doi [\apj]
  {10.1088/0004-637X/795/1/58}, \href
  {https://ui.adsabs.harvard.edu/abs/2014ApJ...795...58L} {795, 58}

\bibitem[\protect\citeauthoryear{{Liu}, {Wang}, {Mao}  \& {Wei}}{{Liu}
  et~al.}{2010}]{2010ApJ...715L.113L}
{Liu} H.,  {Wang} J.,  {Mao} Y.,   {Wei} J.,  2010, \mn@doi [\apjl]
  {10.1088/2041-8205/715/2/L113}, \href
  {https://ui.adsabs.harvard.edu/abs/2010ApJ...715L.113L} {715, L113}

\bibitem[\protect\citeauthoryear{{Mao} \& {Yi}}{{Mao} \&
  {Yi}}{2021}]{2021ApJS..255...10M}
{Mao} L.,  {Yi} T.,  2021, \mn@doi [\apjs] {10.3847/1538-4365/abfd3b}, \href
  {https://ui.adsabs.harvard.edu/abs/2021ApJS..255...10M} {255, 10}

\bibitem[\protect\citeauthoryear{{Mundo} et~al.,}{{Mundo}
  et~al.}{2020}]{2020MNRAS.496.2922M}
{Mundo} S.~A.,  et~al., 2020, \mn@doi [\mnras] {10.1093/mnras/staa1744}, \href
  {https://ui.adsabs.harvard.edu/abs/2020MNRAS.496.2922M} {496, 2922}

\bibitem[\protect\citeauthoryear{{Ojha}, {Jha}, {Chand}  \& {Singh}}{{Ojha}
  et~al.}{2022}]{2022MNRAS.tmp.1571O}
{Ojha} V.,  {Jha} V.~K.,  {Chand} H.,   {Singh} V.,  2022, \mn@doi [\mnras]
  {10.1093/mnras/stac1627}, \href
  {https://ui.adsabs.harvard.edu/abs/2022MNRAS.tmp.1571O} {}

\bibitem[\protect\citeauthoryear{{Olgu{\'\i}n-Iglesias}, {Kotilainen}  \&
  {Chavushyan}}{{Olgu{\'\i}n-Iglesias} et~al.}{2020}]{2020MNRAS.492.1450O}
{Olgu{\'\i}n-Iglesias} A.,  {Kotilainen} J.,   {Chavushyan} V.,  2020, \mn@doi
  [\mnras] {10.1093/mnras/stz3549}, \href
  {https://ui.adsabs.harvard.edu/abs/2020MNRAS.492.1450O} {492, 1450}

\bibitem[\protect\citeauthoryear{{Oshlack}, {Webster}  \& {Whiting}}{{Oshlack}
  et~al.}{2001}]{2001ApJ...558..578O}
{Oshlack} A.~Y.~K.~N.,  {Webster} R.~L.,   {Whiting} M.~T.,  2001, \mn@doi
  [\apj] {10.1086/322299}, \href
  {https://ui.adsabs.harvard.edu/abs/2001ApJ...558..578O} {558, 578}

\bibitem[\protect\citeauthoryear{{Paliya} \& {Stalin}}{{Paliya} \&
  {Stalin}}{2016}]{2016ApJ...820...52P}
{Paliya} V.~S.,  {Stalin} C.~S.,  2016, \mn@doi [\apj]
  {10.3847/0004-637X/820/1/52}, \href
  {https://ui.adsabs.harvard.edu/abs/2016ApJ...820...52P} {820, 52}

\bibitem[\protect\citeauthoryear{{Paliya}, {Ajello}, {Rakshit}, {Mandal},
  {Stalin}, {Kaur}  \& {Hartmann}}{{Paliya} et~al.}{2018}]{2018ApJ...853L...2P}
{Paliya} V.~S.,  {Ajello} M.,  {Rakshit} S.,  {Mandal} A.~K.,  {Stalin} C.~S.,
  {Kaur} A.,   {Hartmann} D.,  2018, \mn@doi [\apjl]
  {10.3847/2041-8213/aaa5ab}, \href
  {https://ui.adsabs.harvard.edu/abs/2018ApJ...853L...2P} {853, L2}

\bibitem[\protect\citeauthoryear{{Pan}, {Yuan}, {Yao}, {Komossa}  \&
  {Jin}}{{Pan} et~al.}{2018}]{2018ApJ...866...69P}
{Pan} H.-W.,  {Yuan} W.,  {Yao} S.,  {Komossa} S.,   {Jin} C.,  2018, \mn@doi
  [\apj] {10.3847/1538-4357/aadd4a}, \href
  {https://ui.adsabs.harvard.edu/abs/2018ApJ...866...69P} {866, 69}

\bibitem[\protect\citeauthoryear{{Poole} et~al.,}{{Poole}
  et~al.}{2008}]{2008MNRAS.383..627P}
{Poole} T.~S.,  et~al., 2008, \mn@doi [\mnras]
  {10.1111/j.1365-2966.2007.12563.x}, \href
  {https://ui.adsabs.harvard.edu/abs/2008MNRAS.383..627P} {383, 627}

\bibitem[\protect\citeauthoryear{{Puchnarewicz} et~al.,}{{Puchnarewicz}
  et~al.}{1992}]{Puchnarewicz1992}
{Puchnarewicz} E.~M.,  et~al., 1992, \mn@doi [\mnras]
  {10.1093/mnras/256.3.589}, \href
  {https://ui.adsabs.harvard.edu/abs/1992MNRAS.256..589P} {256, 589}

\bibitem[\protect\citeauthoryear{{Rakshit}, {Schramm}, {Stalin}, {Tanaka},
  {Paliya}, {Pal}, {Kotilainen}  \& {Shin}}{{Rakshit}
  et~al.}{2021}]{2021MNRAS.504L..22R}
{Rakshit} S.,  {Schramm} M.,  {Stalin} C.~S.,  {Tanaka} I.,  {Paliya} V.~S.,
  {Pal} I.,  {Kotilainen} J.,   {Shin} J.,  2021, \mn@doi [\mnras]
  {10.1093/mnrasl/slab031}, \href
  {https://ui.adsabs.harvard.edu/abs/2021MNRAS.504L..22R} {504, L22}

\bibitem[\protect\citeauthoryear{{Richards} et~al.,}{{Richards}
  et~al.}{2011}]{2011ApJS..194...29R}
{Richards} J.~L.,  et~al., 2011, \mn@doi [\apjs] {10.1088/0067-0049/194/2/29},
  \href {https://ui.adsabs.harvard.edu/abs/2011ApJS..194...29R} {194, 29}

\bibitem[\protect\citeauthoryear{{Richards}, {Hovatta}, {Max-Moerbeck},
  {Pavlidou}, {Pearson}  \& {Readhead}}{{Richards}
  et~al.}{2014}]{2014MNRAS.438.3058R}
{Richards} J.~L.,  {Hovatta} T.,  {Max-Moerbeck} W.,  {Pavlidou} V.,  {Pearson}
  T.~J.,   {Readhead} A.~C.~S.,  2014, \mn@doi [\mnras]
  {10.1093/mnras/stt2412}, \href
  {https://ui.adsabs.harvard.edu/abs/2014MNRAS.438.3058R} {438, 3058}

\bibitem[\protect\citeauthoryear{{Roming} et~al.,}{{Roming}
  et~al.}{2005}]{2005SSRv..120...95R}
{Roming} P. W.~A.,  et~al., 2005, \mn@doi [\ssr] {10.1007/s11214-005-5095-4},
  \href {https://ui.adsabs.harvard.edu/abs/2005SSRv..120...95R} {120, 95}

\bibitem[\protect\citeauthoryear{{Schlafly} \& {Finkbeiner}}{{Schlafly} \&
  {Finkbeiner}}{2011}]{2011ApJ...737..103S}
{Schlafly} E.~F.,  {Finkbeiner} D.~P.,  2011, \mn@doi [\apj]
  {10.1088/0004-637X/737/2/103}, \href
  {https://ui.adsabs.harvard.edu/abs/2011ApJ...737..103S} {737, 103}

\bibitem[\protect\citeauthoryear{{Urry} \& {Padovani}}{{Urry} \&
  {Padovani}}{1995}]{1995PASP..107..803U}
{Urry} C.~M.,  {Padovani} P.,  1995, \mn@doi [\pasp] {10.1086/133630}, \href
  {https://ui.adsabs.harvard.edu/abs/1995PASP..107..803U} {107, 803}

\bibitem[\protect\citeauthoryear{{Wagner}, {Bicknell}  \& {Umemura}}{{Wagner}
  et~al.}{2012}]{Wagner2012}
{Wagner} A.~Y.,  {Bicknell} G.~V.,   {Umemura} M.,  2012, \mn@doi [\apj]
  {10.1088/0004-637X/757/2/136}, \href
  {https://ui.adsabs.harvard.edu/abs/2012ApJ...757..136W} {757, 136}

\bibitem[\protect\citeauthoryear{{Wang} et~al.,}{{Wang}
  et~al.}{2016}]{2016ApJ...824..149W}
{Wang} F.,  et~al., 2016, \mn@doi [\apj] {10.3847/0004-637X/824/2/149}, \href
  {https://ui.adsabs.harvard.edu/abs/2016ApJ...824..149W} {824, 149}

\bibitem[\protect\citeauthoryear{{White} \& {Becker}}{{White} \&
  {Becker}}{1992}]{1992ApJS...79..331W}
{White} R.~L.,  {Becker} R.~H.,  1992, \mn@doi [\apjs] {10.1086/191656}, \href
  {https://ui.adsabs.harvard.edu/abs/1992ApJS...79..331W} {79, 331}

\bibitem[\protect\citeauthoryear{{Wright} \& {Otrupcek}}{{Wright} \&
  {Otrupcek}}{1990}]{1990PKS...C......0W}
{Wright} A.,  {Otrupcek} R.,  1990, PKS Catalog (1990, \href
  {https://ui.adsabs.harvard.edu/abs/1990PKS...C......0W} {p.~0}

\bibitem[\protect\citeauthoryear{{Xin}, {Xiong}, {Bai}, {Liu}, {Lu}  \&
  {Mao}}{{Xin} et~al.}{2022}]{2022RAA....22g5001X}
{Xin} Y.-X.,  {Xiong} D.-R.,  {Bai} J.-M.,  {Liu} H.-T.,  {Lu} K.-X.,   {Mao}
  J.-R.,  2022, \mn@doi [Research in Astronomy and Astrophysics]
  {10.1088/1674-4527/ac684e}, \href
  {https://ui.adsabs.harvard.edu/abs/2022RAA....22g5001X} {22, 075001}

\bibitem[\protect\citeauthoryear{{Yang} et~al.,}{{Yang}
  et~al.}{2018}]{2018MNRAS.477.5127Y}
{Yang} H.,  et~al., 2018, \mn@doi [\mnras] {10.1093/mnras/sty904}, \href
  {https://ui.adsabs.harvard.edu/abs/2018MNRAS.477.5127Y} {477, 5127}

\bibitem[\protect\citeauthoryear{{Yao}, {Yuan}, {Komossa}, {Grupe}, {Fuhrmann}
  \& {Liu}}{{Yao} et~al.}{2015a}]{2015AJ....150...23Y}
{Yao} S.,  {Yuan} W.,  {Komossa} S.,  {Grupe} D.,  {Fuhrmann} L.,   {Liu} B.,
  2015a, \mn@doi [\aj] {10.1088/0004-6256/150/1/23}, \href
  {https://ui.adsabs.harvard.edu/abs/2015AJ....150...23Y} {150, 23}

\bibitem[\protect\citeauthoryear{{Yao}, {Yuan}, {Zhou}, {Komossa}, {Zhang},
  {Qiao}  \& {Liu}}{{Yao} et~al.}{2015b}]{2015MNRAS.454L..16Y}
{Yao} S.,  {Yuan} W.,  {Zhou} H.,  {Komossa} S.,  {Zhang} J.,  {Qiao} E.,
  {Liu} B.,  2015b, \mn@doi [\mnras] {10.1093/mnrasl/slv119}, \href
  {https://ui.adsabs.harvard.edu/abs/2015MNRAS.454L..16Y} {454, L16}

\bibitem[\protect\citeauthoryear{{Yao}, {Komossa}, {Liu}, {Yi}, {Yuan}, {Zhou}
  \& {Wu}}{{Yao} et~al.}{2019}]{2019MNRAS.487L..40Y}
{Yao} S.,  {Komossa} S.,  {Liu} W.-J.,  {Yi} W.,  {Yuan} W.,  {Zhou} H.,   {Wu}
  X.-B.,  2019, \mn@doi [\mnras] {10.1093/mnrasl/slz071}, \href
  {https://ui.adsabs.harvard.edu/abs/2019MNRAS.487L..40Y} {487, L40}

\bibitem[\protect\citeauthoryear{{Yuan}, {Zhou}, {Komossa}, {Dong}, {Wang},
  {Lu}  \& {Bai}}{{Yuan} et~al.}{2008}]{2008ApJ...685..801Y}
{Yuan} W.,  {Zhou} H.~Y.,  {Komossa} S.,  {Dong} X.~B.,  {Wang} T.~G.,  {Lu}
  H.~L.,   {Bai} J.~M.,  2008, \mn@doi [\apj] {10.1086/591046}, \href
  {https://ui.adsabs.harvard.edu/abs/2008ApJ...685..801Y} {685, 801}

\bibitem[\protect\citeauthoryear{{Zhou}, {Wang}, {Dong}, {Zhou}  \&
  {Li}}{{Zhou} et~al.}{2003}]{2003ApJ...584..147Z}
{Zhou} H.-Y.,  {Wang} T.-G.,  {Dong} X.-B.,  {Zhou} Y.-Y.,   {Li} C.,  2003,
  \mn@doi [\apj] {10.1086/345523}, \href
  {https://ui.adsabs.harvard.edu/abs/2003ApJ...584..147Z} {584, 147}

\bibitem[\protect\citeauthoryear{{Zhou} et~al.,}{{Zhou}
  et~al.}{2007}]{2007ApJ...658L..13Z}
{Zhou} H.,  et~al., 2007, \mn@doi [\apjl] {10.1086/513604}, \href
  {https://ui.adsabs.harvard.edu/abs/2007ApJ...658L..13Z} {658, L13}

\makeatother
\end{thebibliography}




\appendix


\bsp	
\label{lastpage}
\end{document}